\documentclass[12pt]{article}

\usepackage{graphicx}
\usepackage{amssymb}
\usepackage{amsmath}
\usepackage{amssymb}
\usepackage{amsthm}
\usepackage{amscd}
\usepackage{amsfonts}
\usepackage{caption}
 \usepackage{epstopdf}
 \usepackage[table]{xcolor}
\usepackage{fancyhdr}
\usepackage{float}

\usepackage[T1]{fontenc} 

\usepackage{amsthm}
\usepackage{amsopn}

\usepackage{multirow}
\usepackage{subcaption}
\usepackage[legalpaper,margin=1.5in]{geometry}

\begin{document}
\title{Kink Dynamics in a Nonlinear Beam Model}

\author{Robert J.\ Decker$^{1}$, \;\;  A.\ Demirkaya$^{1}$,  \;\;P.~G.\ Kevrekidis $^{2}$\\
\;\; Digno Iglesias$^{1}$,  \;\;Jeff Severino$^{1}$,  \;\; Yonathan Shavit$^{1}$\\
  $^{1}$ University of Hartford, \\ $^{2}$ Department of Mathematics
  \& Statistics, \\
  University of Massachusetts, Amherst 01003 USA}

\maketitle

\begin{abstract}
In this paper, we study the single kink and the kink-antikink collisions of
a nonlinear beam equation bearing a fourth-derivative term.
We numerically explore some of the key characteristics of
the single kink both in its standing wave and in its traveling wave form.
A point of emphasis is the study of kink-antikink collisions,
exploring
the critical velocity for single-bounce (and separation) and
infinite-bounce (where the kink and antikink trap each other)
windows.
The relevant phenomenology turns out to be dramatically different than
that
of the corresponding nonlinear Klein-Gordon (i.e., $\phi^4$) model. 
Our computations show that for small initial velocities, the kink and
antikink reflect nearly elastically without colliding. For an intermediate interval of velocities, the two waves trap each
other, while for large speeds a single inelastic collision between
them takes place.
Lastly, we
briefly touch upon the use of collective coordinates (CC) method and
their
predictions of the relevant phenomenology. When one degree of freedom
is used in the CC approach, the results match well the numerical ones
for small values of initial velocity. However, for bigger values of
initial velocity,
it is inferred that more degrees of freedom need to be
self-consistently included in order to capture the collision phenomenology.\end{abstract}

\section{Introduction}
Different variants of the
nonlinear beam equation has been studied in the last decade both
numerically and analytically; see, e.g.,
\cite{levandosky,champneys,CM,karageorgis}.
Such models have been been considered chiefly in the context of
suspension
bridges and the propagation of traveling waves therein (most notably
for
piecewise constant but also for exponential nonlinearities); see
the relevant discussion in~\cite{champneys,CM,karageorgis}.
More recently, different venues of interest of such fourth-derivative
settings
have arisen both at the level of applications where they have emerged
in generalized nonlinear Schr{\"o}dinger (NLS) settings involving so-called
pure-quartic solitons in nonlinear optics~\cite{pqs}, but also
equally importantly in the realm of mathematical analysis in
connection
to their intriguing existence and stability properties~\cite{atanas}.

One of the particularly intriguing aspects of this class of models
is that  the standing and traveling
waves of the beam equation satisfy a fourth-order
ordinary differential equation, whereas for other dispersive
wave models, such as the
Korteweg-de Vries equation,
traveling waves satisfy a
second-order ordinary differential equation. The same is naturally
true for well established models such as the standard NLS
equation
and the Klein-Gordon family of models~\cite{ablowitz}. 
Since there is no explicit formula for the standing and traveling waves, it is challenging to obtain the spectral information analytically.  In \cite{levandosky}, the existence of ground-state solitary traveling wave solutions 
was shown by using a constrained minimization technique.
The corresponding Hessian was used to infer stability information in
that work; e.g., traveling waves were found to be stable at least in
the vicinity of a critical value for power law nonlinearities of
sufficiently
low power. At the same time, standing waves (for low enough
nonlinearity powers) were found to be stable for a suitably frequency
interval.
In \cite{beam_demirkaya},  the existence and the stability of standing
and traveling waves
for the same setting as that of~\cite{levandosky} was studied
numerically for a number of one-dimensional case examples.
The authors of \cite{karageorgis} showed the existence of traveling
wave solutions for a large class of nonlinearities by adapting the
Nehari manifold approach; this approach, however, does not provide
information for the stability of the waves. 

\begin{figure}[h]
\begin{center}
\includegraphics[width=7cm, height=2cm]{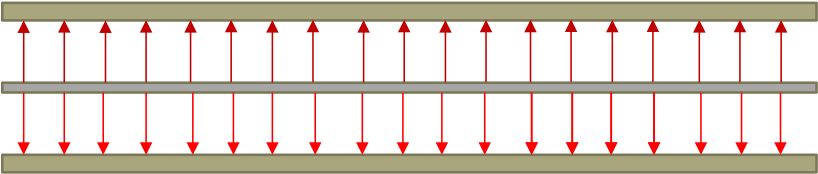}
\end{center}
\caption{A schematic of a thin magnetic metal beam suspended between two electromagnets at $u=\pm 1$.}
\label{beam_f}
\end{figure}

{In this paper, we numerically explore the existence and the behavior of kink and kink-antikink solutions of a nonlinear beam equation: 
\begin{equation}
u_{tt}=-u_{xxxx}-V^{\prime }(u)  \label{beam}
\end{equation}%
where $V(u)=\frac{1}{2}(u^{2}-1)^{2}$. This potential function is a
departure from the papers described in the previous paragraphs. In
particular, it represents a double-well potential, and therefore admits
possible kink-antikink (topological soliton) solutions. For example \cite%
{levandosky}, \cite{karageorgis} and \cite{beam_demirkaya} address potential
functions that include $V(u)=-\frac{1}{2}(u^{2}-1)^{2}$ (and generalizations
thereof) which makes $u=\pm 1$ unstable and $u=0$ stable (the opposite of
ours). We have chosen our potential function so that we can make comparisons
with the well studied $\phi ^{4}$ model, where $u_{xxxx}$ in our model is
replaced by $-u_{xx}$.

We are not aware of any definitive previous proposals of
  a physical setting described by the above model.
However, we suggest that a thin magnetic metal beam
suspended between two electromagnets as shown in Figure \ref{beam_f} could
represent a reasonable physical system with the same properties as our PDE
model, in the same way that the $\phi ^{4}$ equation would reasonably
correspond to a thin (very) flexible magnetic metal wire suspended between
two electromagnets. Similar reasoning has been used with the well-known
Duffing equation (ODE); in \cite{Moon} the authors report
on creating a realistic physical model of a flexible beam suspended between two magnets, which is compared favorably to the
predicted theory; the potential function is the same as the one we use.
Also, our interpretations are closely related to the classical
interpretations of the linear wave equation ($u_{tt}=u_{xx}$) and the linear
beam equation ($u_{tt}=-u_{xxxx}$) as representing small vibrations of a
flexible string and a beam respectively.}


The nonlinear beam model (\ref{beam}) is similar to the $\phi^4$ model
\begin{equation}\label{phi4} 
u_{tt}=u_{xx}-V'(u)
\end{equation}
 which has been studied intensely both analytically and numerically
 over three decades now \cite{Campbell,Belova}; see also the recent
 book~\cite{cuevas}
 summarizing the current state of understanding for such Klein-Gordon
 models. 
 Our aim in the present first work is to present some of the basic
 features of the biharmonic analogue of the $\phi^4$ model, which we
 will
 hereafter term biharmonic $\phi^4$ or B$\phi^4$ for short. 
In the present work, we first present numerical computations and
simulations for a single kink at the level of both standing and
traveling waves.  Next, we study the behavior of kink-antikink
solutions which is well-known to be particularly elaborate in
the
standard $\phi^4$
model~\cite{Campbell,Belova,Ann,goodman,goodman2,weigel,weigel2}.
The latter, per the recent work of~\cite{weigel,weigel2} (see
also~\cite{cuevas}) is still an ongoing research theme.
Here, we show that the interactions between kink and antikink are in
some ways much simpler, yet at the same time in other ways
fundamentally
more complex. The interactions up to speeds of the incoming wave of
about $0.5$ are nearly elastic and, importantly, {\it effectively repulsive},
i.e., the kink and antikink never get to reach the same location while
interacting. For large speeds between $0.6$ and $1$ the
large kinetic energy of the coherent structures overcomes their
interaction barrier and leads to collision and separation with the waves
moving
at speeds lower than the
incoming ones. In between, a delicate trapping window arises with edges
featuring a very complex (oscillatory and logarithmic) dependence of
the
outgoing vs. the incoming velocity.
We present the relevant dependencies, for the first time to our
knowledge, and expose some of the interesting questions arising from
our numerical computations worthwhile to address in future studies.

\section{Numerical Methods}

In order to simulate Eq. (\ref{beam}) numerically we discretize the
spatial domain on the interval $x\in[-100, 100]$ with an increment of
$\Delta x=0.1$. We use a Fourier-based spectral
differentiation matrix $D_2$ as in \cite{trefethen} to
approximate $\varphi''$ as $D_2\varphi$  and to approximate
$\varphi^{(4)}$ as $D_2^{2}\varphi$.  This turns the PDE (\ref{beam})
into a system of ODE's and we use Matlab's built in ODE solver
\textit{ode45} to simulate the kink and antikink evolution therein.

\section{Single Kink Solutions}
A kink solution for Eq. (\ref{beam}) (or for Eq. (\ref{phi4})) is a
solution for which $u\rightarrow\pm1$ as $x\rightarrow\pm\infty$
respectively, as shown in Figure \ref{steady} in the first panel. An
antikink is a solution for which $u\rightarrow\mp1$ as
$x\rightarrow\pm\infty$ respectively; an antikink can be obtained from
a kink by reflection about either the horizontal or vertical axis. In
this section, we study the behavior of a single kink solution
numerically. We start with the steady state solution, and study its
existence and stability numerically. Next, we consider the moving
single kink solutions,
examining their corresponding properties.
We also briefly touch upon
energy and momentum conservation considerations
indicating the corresponding properties of the models and examining
them as a numerical
check the validity of our direct simulations.


\subsection{Steady state Kink Solutions}
\begin{figure}[H]
\begin{center}
\includegraphics[width=0.46\textwidth]{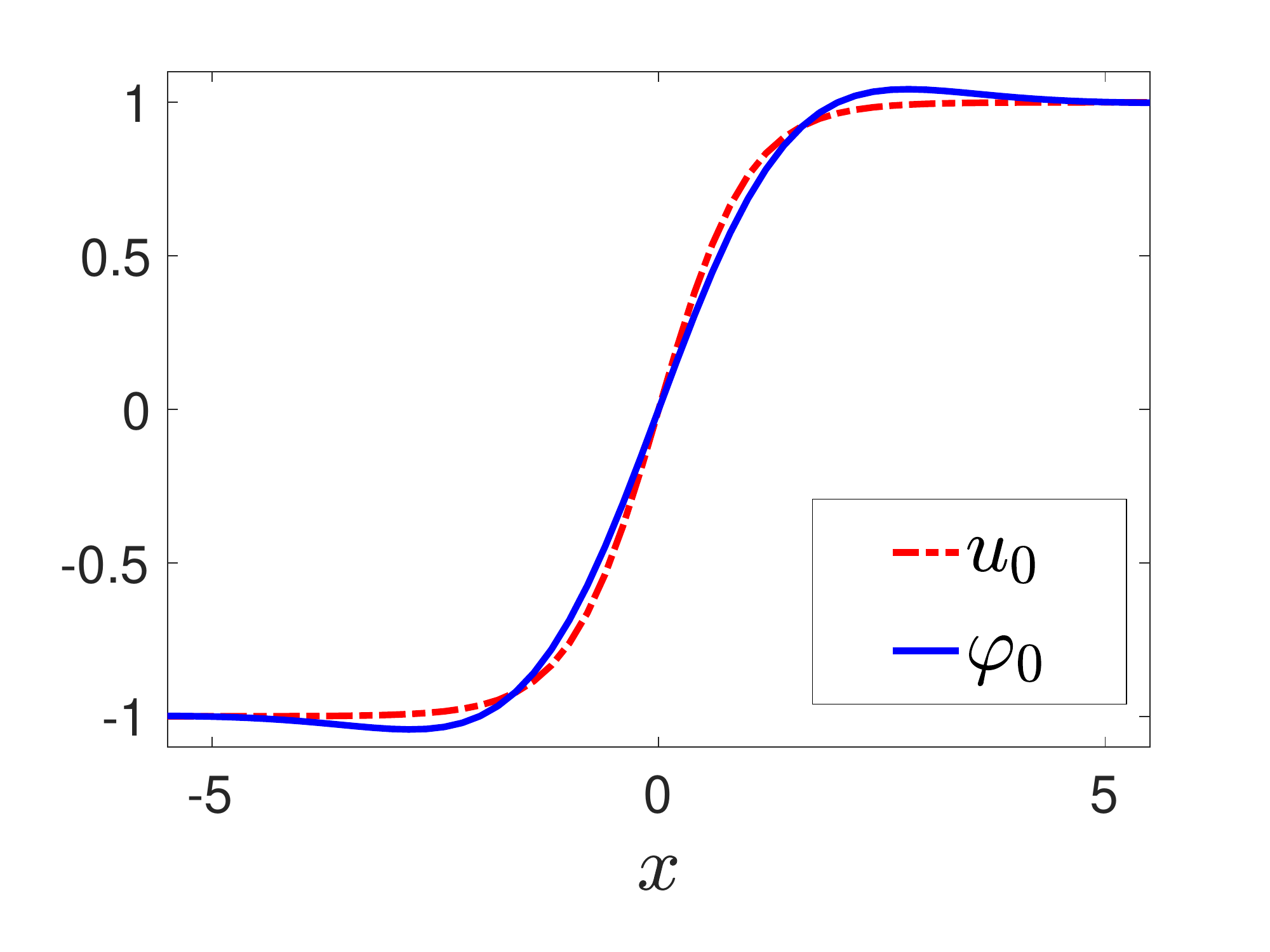}
\includegraphics[width=0.46\textwidth]{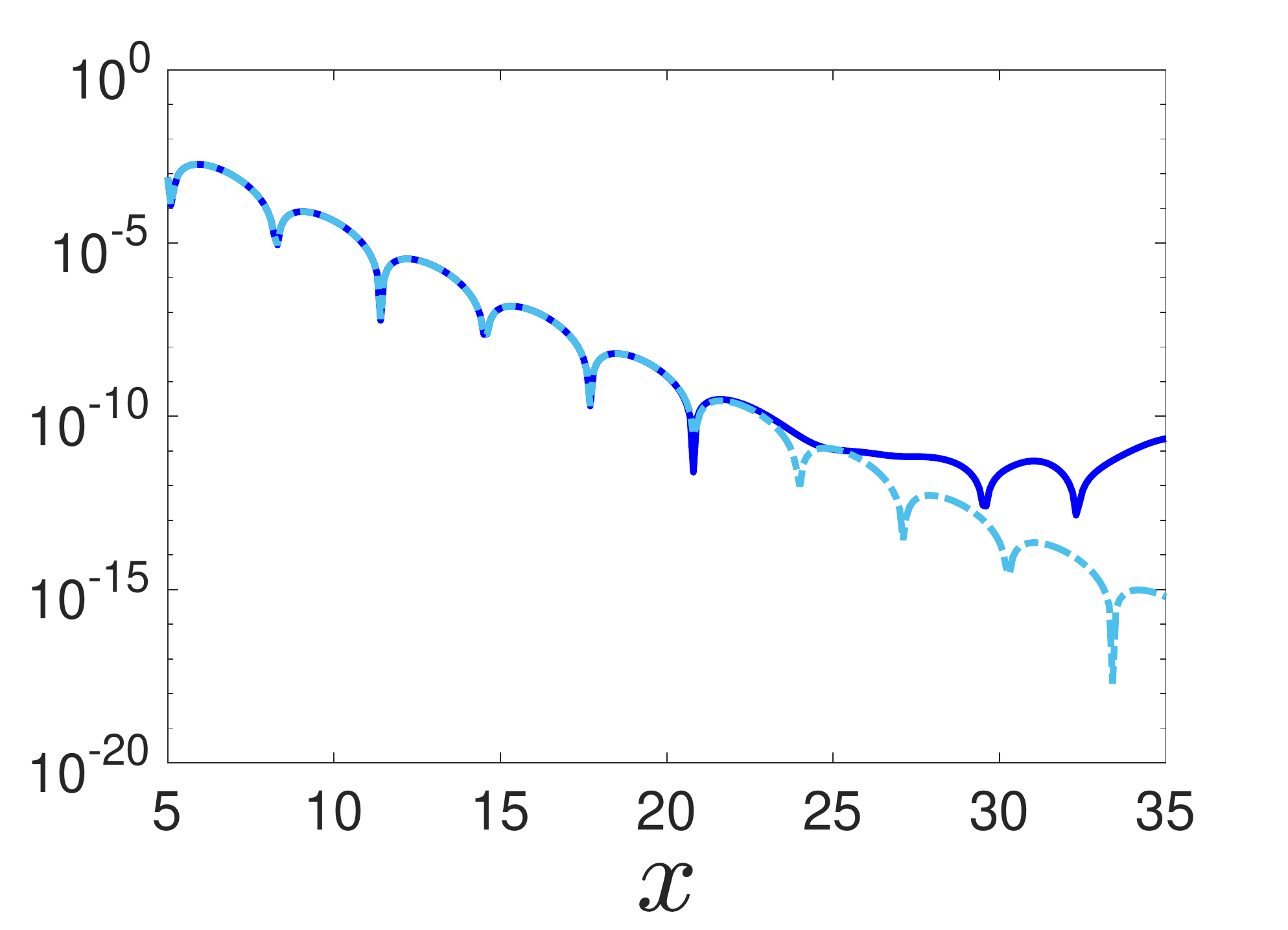}
\includegraphics[width=0.46\textwidth]{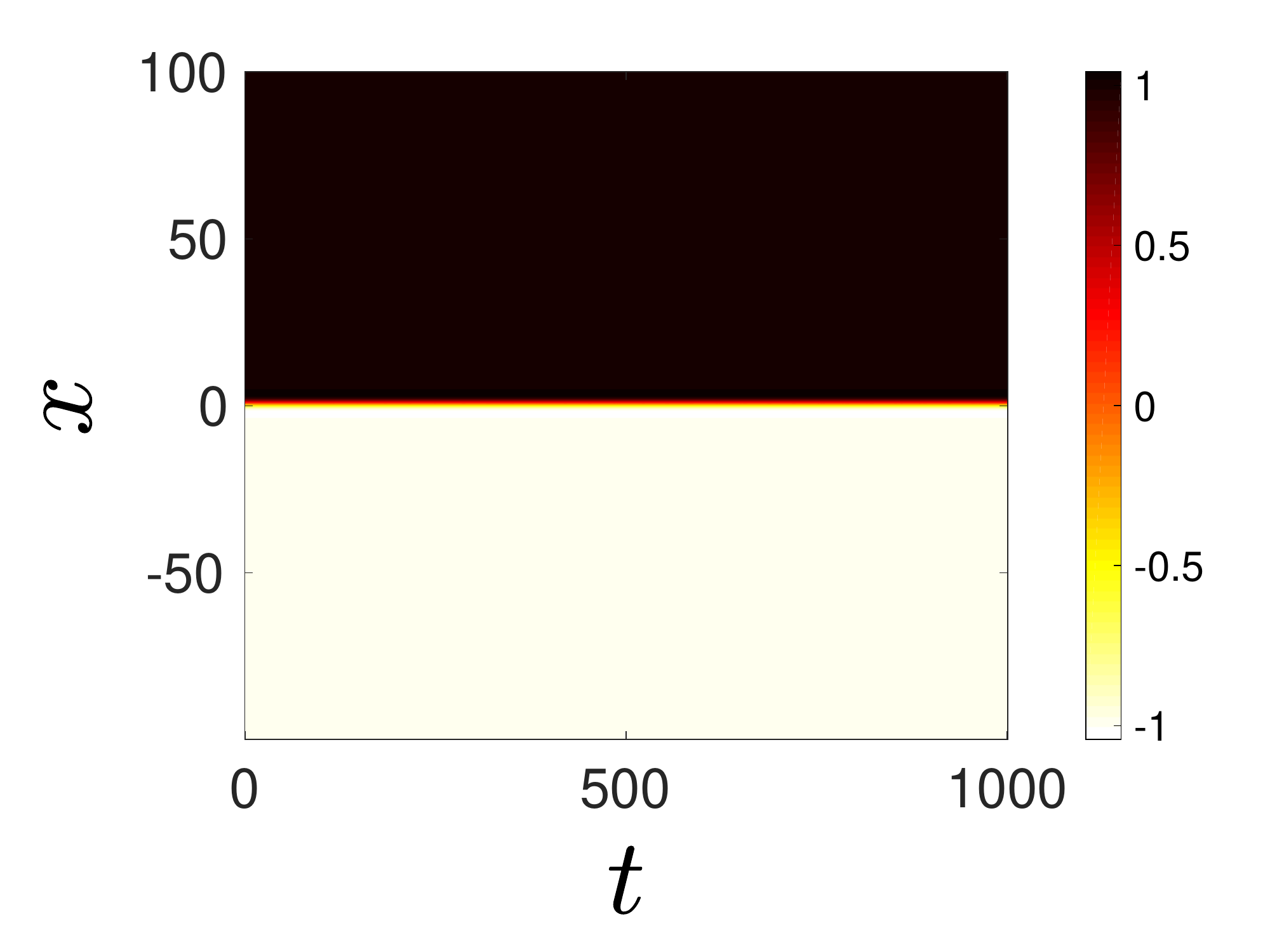}
\end{center}

\caption{The top left panel shows the steady state kinks for the
  B$\phi^4$ model (blue solid line) and $\phi^4$ model (red dashed
  line).
Notice the oscillatory nature of the former in comparison with the
monotonic nature of the latter.
  The top right panel shows the curves $|\varphi_0-1|$ (blue solid line)
  and the fitted curve  $e^{-0.9998x} |(0.965\cos(0.9998(x-
  0.4086)))|$ (light blue dash-dotted line). The bottom panel is the
  space-time (i.e., $x-t$) contour plot of the (dynamically robust)
  steady kink evolution. }
\label{steady}
\end{figure}

Steady state kink solutions $u(x,t)=\varphi_{0}(x)$ of Eq. (\ref{beam}) satisfy
\begin{equation}\label{beam_steady}
\varphi_{0}^{(4)}+V'(\varphi_{0})=0.
\end{equation} 
We numerically solve this fourth order BVP using Matlab's
\textit{fsolve} and choose as
initial guess the explicitly known solution to the steady-state
$\phi^4$ model, namely, $u_0(x)=\tanh(x)$. The result of the
corresponding
computation is shown in the top left panel of Fig.~\ref{steady}.

It is worthwhile to briefly consider the asymptotics of the relevant
kink, i.e., how it approaches the homogeneous steady states $u=\pm 1$.
Substituting $\varphi_0(x)=1-\epsilon e^{\lambda x}$ into Eq. \ref{beam_steady},  we obtain
(as $\epsilon \to 0$) $\lambda^4+4 = 0$; choosing the root
$\lambda=-1+i$,
we get $\varphi_0(x) \approx 1-\epsilon e^{-x} \cos(x-x_0)$ for small
$\epsilon$, where $x_0$ denotes a suitable constant. 
In Fig. \ref{steady}, the top right panel shows the plots for $|\varphi_0 -1|$ and the fitted curve for the function in the form:
\begin{equation}\label{fitted}
e^{-ax} |(b\cos(c(x-d)))|
\end{equation}
 where $a, b, c$ and $d$ are parameters. We use Matlab's
 \textit{lsqcurvefit} function to find the values and 95\% confidence
 intervals for $a,b,c$ and $d$. The values and the intervals for those
 parameters are presented in Table \ref{tab:table1}. As seen in the
 Table, the numerically obtained intervals support the theory where
 $a$ and $c$ (the exponential spatial decay rate and the wavenumber of
 the spatial oscillation) are
 expected to be 1.  Note that the fit in the top right panel of
 Fig.~\ref{steady}
 is excellent with a divergence
 occurring
 at around $x=22$ due to the accuracy settings used in finding the
 numerical solution $\varphi_0(x)$.
The bottom panel of Fig.~\ref{steady} illustrates the dynamical
evolution
of the relevant coherent structure predisposing us through its robust
dynamical evolution for the spectral stability of the kink to which we
now turn below.

\begin{table}[h!]
  \begin{center}
    \caption{}
    \label{tab:table1}
    \begin{tabular}{c|c|c} 
      \textbf{Parameters} & \textbf{Values} & \textbf{95 \% CI} \\
      \hline
      a & 0.9998 & [0.9978, 1.0019] \\
      b & 0.9650 & [0.9525, 0.9775]  \\
      c & 0.9998 & [0.9984, 1.0013]  \\
      d & 0.4086 & [0.4015, 0.4156]   \\
    \end{tabular}
    
  \end{center}
\end{table}

\begin{figure}[H]
\begin{center}

\includegraphics[width=0.46\textwidth]{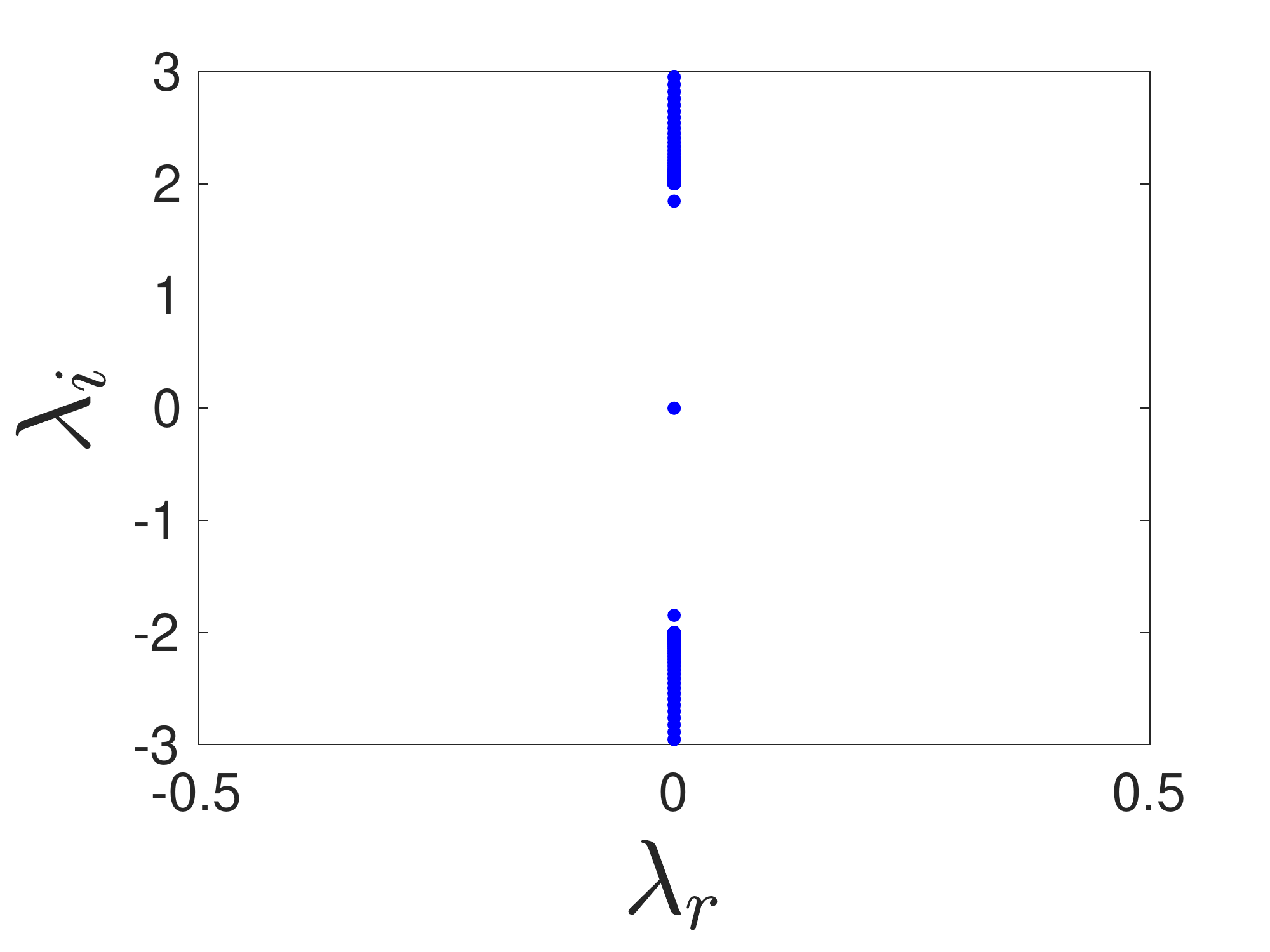}
\includegraphics[width=0.46\textwidth]{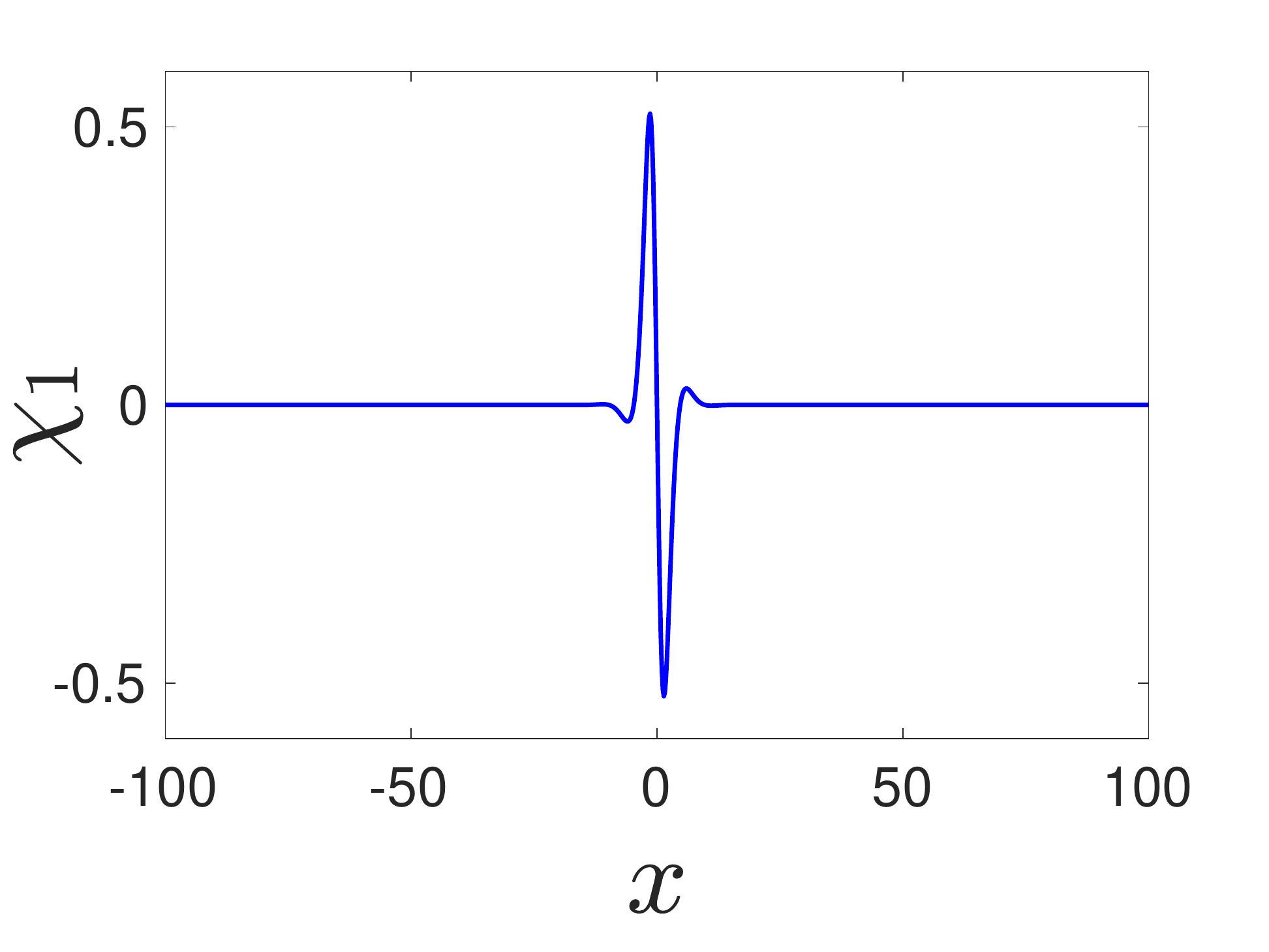}

\end{center}
\caption{The left panel shows the spectral plane $(\lambda_r,
  \lambda_i)$ of linearization eigenvalues $\lambda = \lambda_r +
  i\lambda_i$ corresponding to the steady state of the B$\phi^4$
  model.
  The right panel
  shows the eigenfunction corresponding to the internal mode at
  $\lambda=\pm 1.8458 i$. This is an internal, vibrational anti-symmetric
mode in analogy with the one at $\lambda=\pm \sqrt{3} i$ of the
regular $\phi^4$
model.}
\label{eigvalues}
\end{figure}

To study the stability of the steady state, we consider the linearization around the steady kink solution. Assume 
\begin{equation}\label{steady-perturb}u(x ,t )=\varphi_0 (x)+v (x ,t),
\end{equation}
where $v(x,t)$ is the perturbation assumed to be small when $t=0$. When we substitute Eq. (\ref{steady-perturb}) into Eq.  (\ref{beam}), we get the linearized equation as 
\begin{equation}\label{steady-linear}
v_{tt}=-v_{xxxx}-V''(\varphi_0)v
\end{equation} 
Defining $w (x ,t )=v_t(x ,t)$, we can convert Eq. (\ref{steady-linear}) into a first order linear system 
\begin{equation}
\frac{\partial }{\partial t }\left[ 
\begin{array}{c}
v  \\ 
w
\end{array}%
\right] =\mathcal{L}_{0} \left[ 
\begin{array}{c}
v  \\ 
w
\end{array}%
\right], 
\end{equation}
where 

\begin{equation}
\mathcal{L}_0=\left[ 
\begin{array}{cc}
0 & I \\ 
-D_{xxxx }-V^{\prime \prime }(\varphi_0 )I & 
0
\end{array}
\right]. 
\end{equation}
We solve the relevant spectral eigenvalue problem (of the operator
$\mathcal{L}_0$)
numerically.
In Fig. \ref{eigvalues}, we show the eigenvalues $\lambda = \lambda_r
+ i\lambda_i$ of this operator and the eigenfunction corresponding to
the internal mode at $\lambda=\pm 1.8458 i$.  As seen in the figure,
the purely imaginary nature of all the eigenvalues indicates that the steady
state kink solution is spectrally stable.
This is, indeed, in line with our numerical observations of Fig.~\ref{steady}.

\subsection{Moving Single Kink Solutions}
In this section, we examine the dynamical evolution of a single kink
solution in the form: $u(x,t)=\varphi(x-ct)$ where $c$ is the
speed. For second order differential equations like the $\phi^4$
model, we can apply a Lorentz transformation to the steady state kink
solutions and obtain the moving ones. However, this is not
the case
for the B$\phi^4$ equation because it is a fourth order differential
equation.

The equation that a traveling wave must satisfy can be found by assuming $u(x,t)=\varphi(x-ct)$ and substituting into Eq.  (\ref{beam}) to get 
\begin{equation}\label{beam_traveling}
\varphi^{(4)}(\xi)+c^2\varphi''(\xi)+V'(\varphi(\xi))=0.
\end{equation}
where $\xi=x-ct$. Thus we solve Eq. (\ref{beam_traveling}) numerically in order to
identify
a numerically accurate traveling wave profile.
We use $u_c(x)=\tanh(x/\sqrt{1-c^2})$ (the known $\phi^{4}$ traveling
wave solution at $t=0$) as an initial guess for $fsolve$ in order to find the moving kink solutions. 

\begin{figure}[H]
\begin{center}
\includegraphics[width=0.46\textwidth]{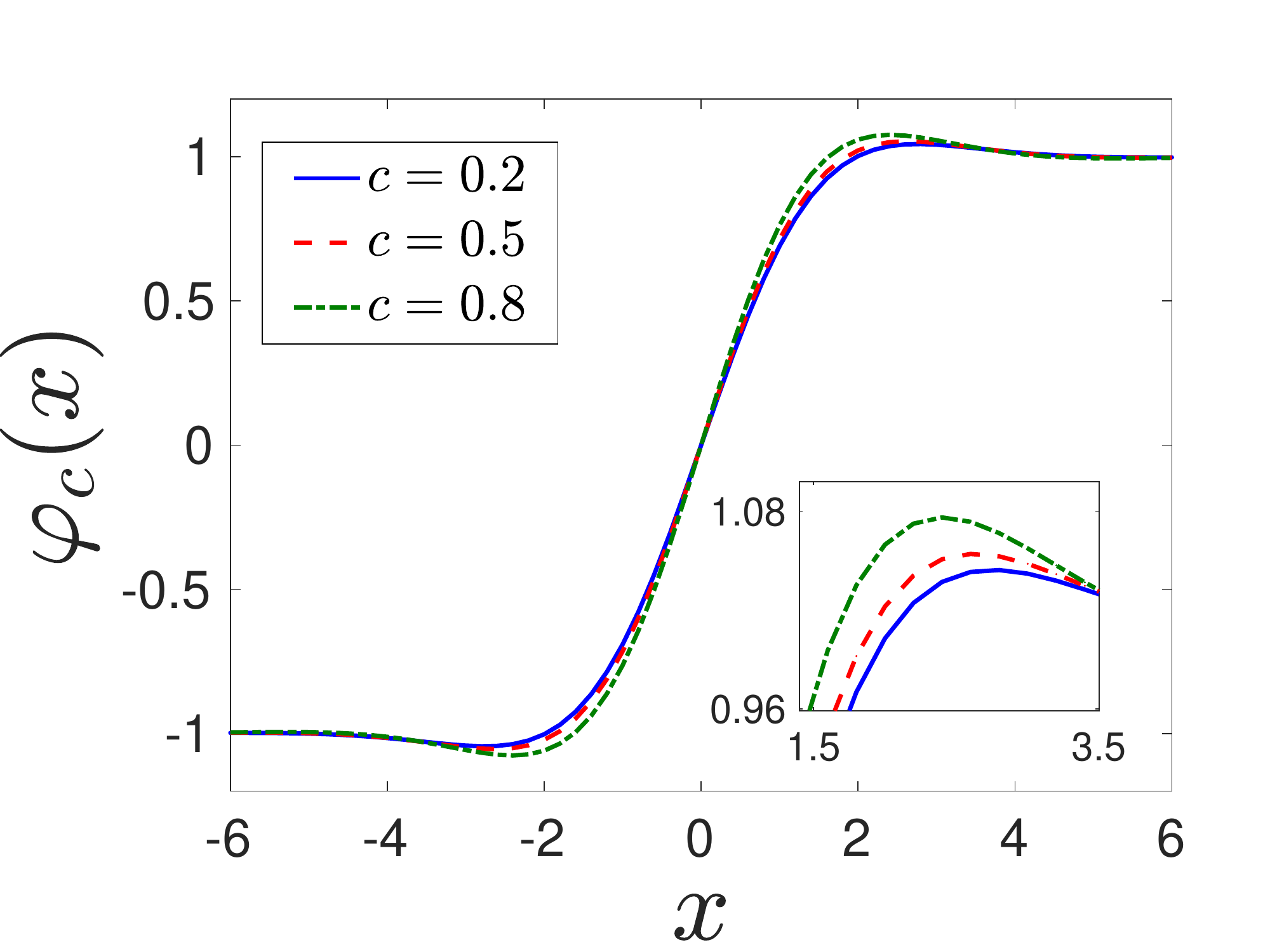}
\includegraphics[width=0.46\textwidth]{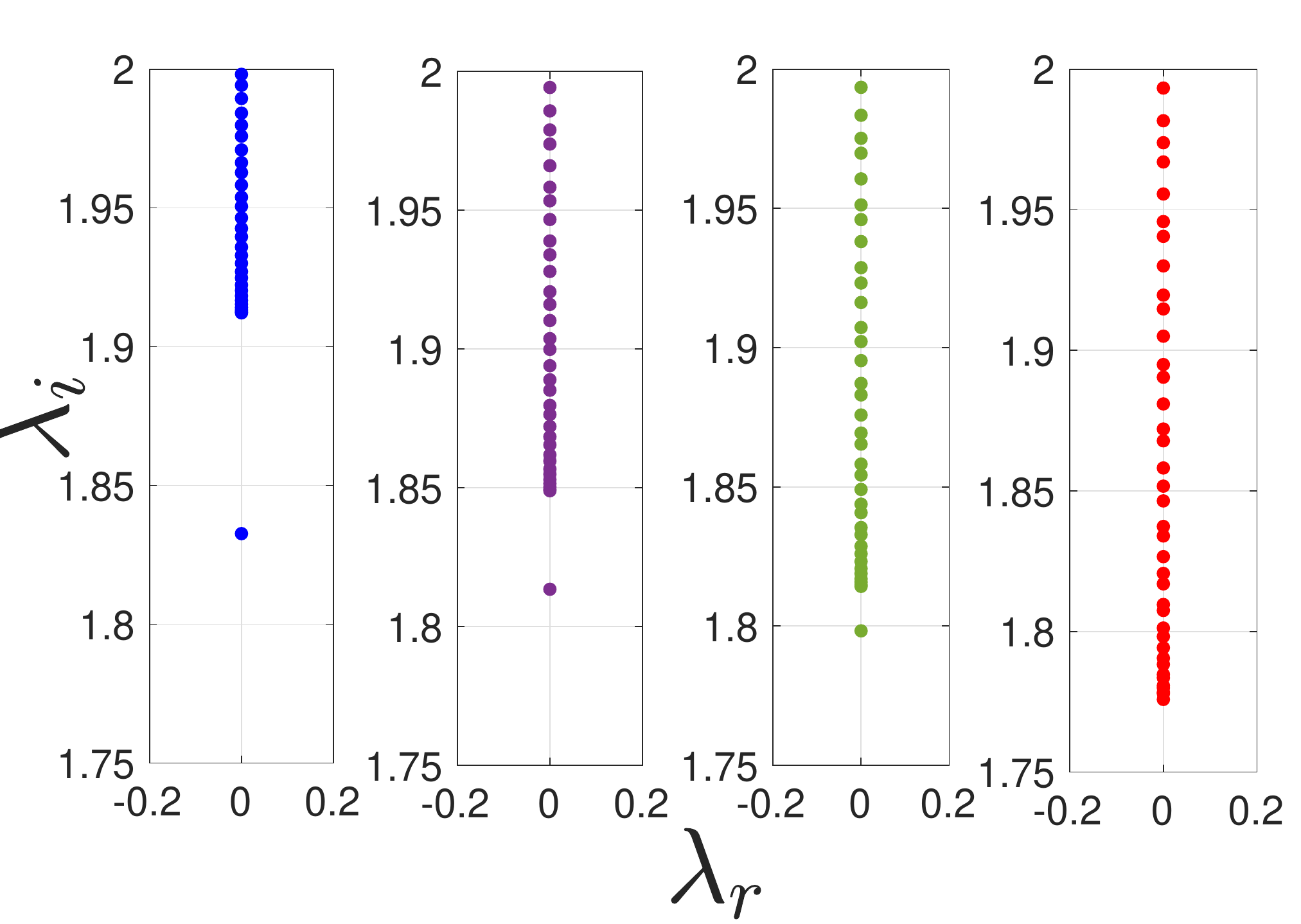} \\
\includegraphics[width=0.46\textwidth]{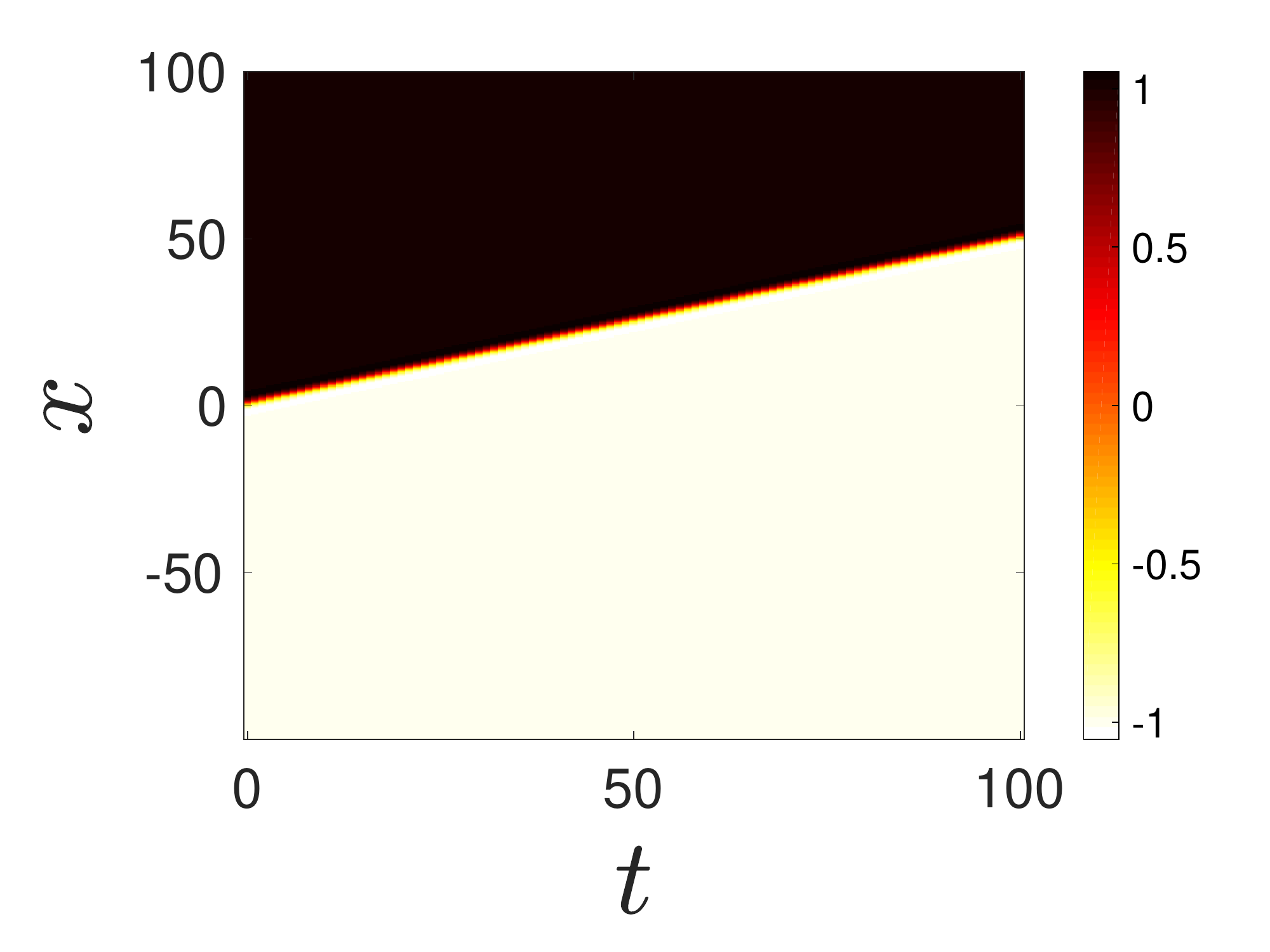}
\end{center}
\caption{The top left panel shows the moving kink solutions for
  $c=0.2$ (blue solid), $c=0.5$ (red dash), $c=0.8$ (green
  dash-dot). The top right panel shows the spectral plane $(\lambda_r,
  \lambda_i)$ of
  the linearization eigenvalues $\lambda = \lambda_r + i\lambda_i$
  associated
  with the moving kink solutions for $c=0.2$, $c=0.3$, $c=0.35$,
  $c=0.4$ from left to right respectively. The bottom panel
  illustrates the contour plot
  of the PDE for speed 
  $c=0.5$ and $x_0=0$.}
\label{movingkink}
\end{figure}

For the stability of these solutions, we study the spectrum of the linearized operator about these moving solutions. Converting Eq. (\ref{beam}) to the new coordinates $\xi =x-ct$ and $\tau =t$ (a moving coordinate system), we obtain 
\begin{equation}\label{newCoords}u_{\tau\tau}=-c^2u_{\xi\xi}+2cu_{\xi\tau}-u_{\xi\xi\xi\xi}-V'(u) \end{equation}
Steady-state solutions of Eq. (\ref{newCoords}) are traveling wave
solutions of Eq. (\ref{beam}) and are given by
Eq. (\ref{beam_traveling}). To
determine stability we assume:
\begin{equation}\label{perturbations}u(\xi ,\tau )=\varphi_c (\xi )+\eta (\xi ,\tau )\end{equation} 
where $\eta$ is the perturbation around the traveling
solution $\varphi_c (\xi )$
and assumed to be small. When we substitute Eq. (\ref{perturbations}) into Eq. (\ref{newCoords}) 
and use Eq. (\ref{beam_traveling}), as well as the approximation $V^{\prime }(\phi_c (\xi )+\eta (\xi ,\tau ))\approx V^{\prime }(\phi_c (\xi
))+\eta (\xi ,\tau )V^{\prime \prime }(\phi_c (\xi ))$, the linearized equation is as follows:
\begin{equation}\label{pert2}
\eta _{\tau \tau }=\left( -c^{2}D_{\xi \xi }-D_{\xi \xi \xi \xi }\right)
\eta (\xi ,\tau )+2cD_{\xi }\eta _{\tau }(\xi ,\tau )-V^{\prime \prime }(\phi_c (\xi ))\eta (\xi ,\tau).
\end{equation}
Defining $\psi (\xi ,\tau )=\eta _{\tau }(\xi ,\tau )$,
Eq. (\ref{pert2}) can
be rewritten as a first order linear system of the form:
\begin{equation}
\frac{\partial }{\partial \tau }\left[ 
\begin{array}{c}
\eta  \\ 
\psi 
\end{array}%
\right] =\mathcal{L}_c \left[ 
\begin{array}{c}
\eta  \\ 
\psi 
\end{array}%
\right], 
\end{equation}
where 

\begin{equation}
\mathcal{L}_c=\left[ 
\begin{array}{cc}
0 & I \\ 
-c^{2}D_{\xi \xi }-D_{\xi \xi \xi \xi }-V^{\prime \prime }(\varphi_c )I & 
2cD_{\xi }
\end{array}
\right]. 
\end{equation}
In Fig. \ref{movingkink}, we show the numerical moving kink solutions
for three values of $c$ and also present the spectra of the linearized
operator $\mathcal{L}_c$ around solutions of different speeds. Importantly, it can be
seen that the relevant solutions are spectrally stable. Additionally,
it can be observed that the first three panels feature an internal mode
with a frequency outside of the continuous spectral band; however, the
rightmost panel associated with speed $c=0.4$ shows no such mode
indicating that apparently the relevant mode has disappeared inside the
continuous
spectrum.

We have also examined the dynamics associated with the relevant
traveling
waves. As a prototypical example, 
by using the initial conditions:
\begin{equation}u(x,0)=\varphi_c(x); \hspace{0.5cm} u_t(x,0)=-c\varphi'_c(x),
\end{equation}
we can simulate a moving single soliton moving with velocity $c$ where
$\varphi_c(x)$ is the solution to Eq. (\ref{beam_traveling}). The
bottom panel in Fig. \ref{movingkink} shows the contour plot of the
moving kink traveling with the speed $c=0.5$.
The relevant solution appears to be robustly propagating for the time
scales considered suggesting that the relevant traveling wave kink is
a genuine stable traveling solution of the original problem of
Eq.~(\ref{beam}).
We have indeed confirmed that similar results can be obtained for
other
speeds, in line with our theoretical analysis (data not shown here).

\subsection{Conservation Laws and Numerical Method Validation}
\subsubsection{Conservation of Energy}

\begin{figure}[H]
\begin{center}
\includegraphics[width=0.46\textwidth]{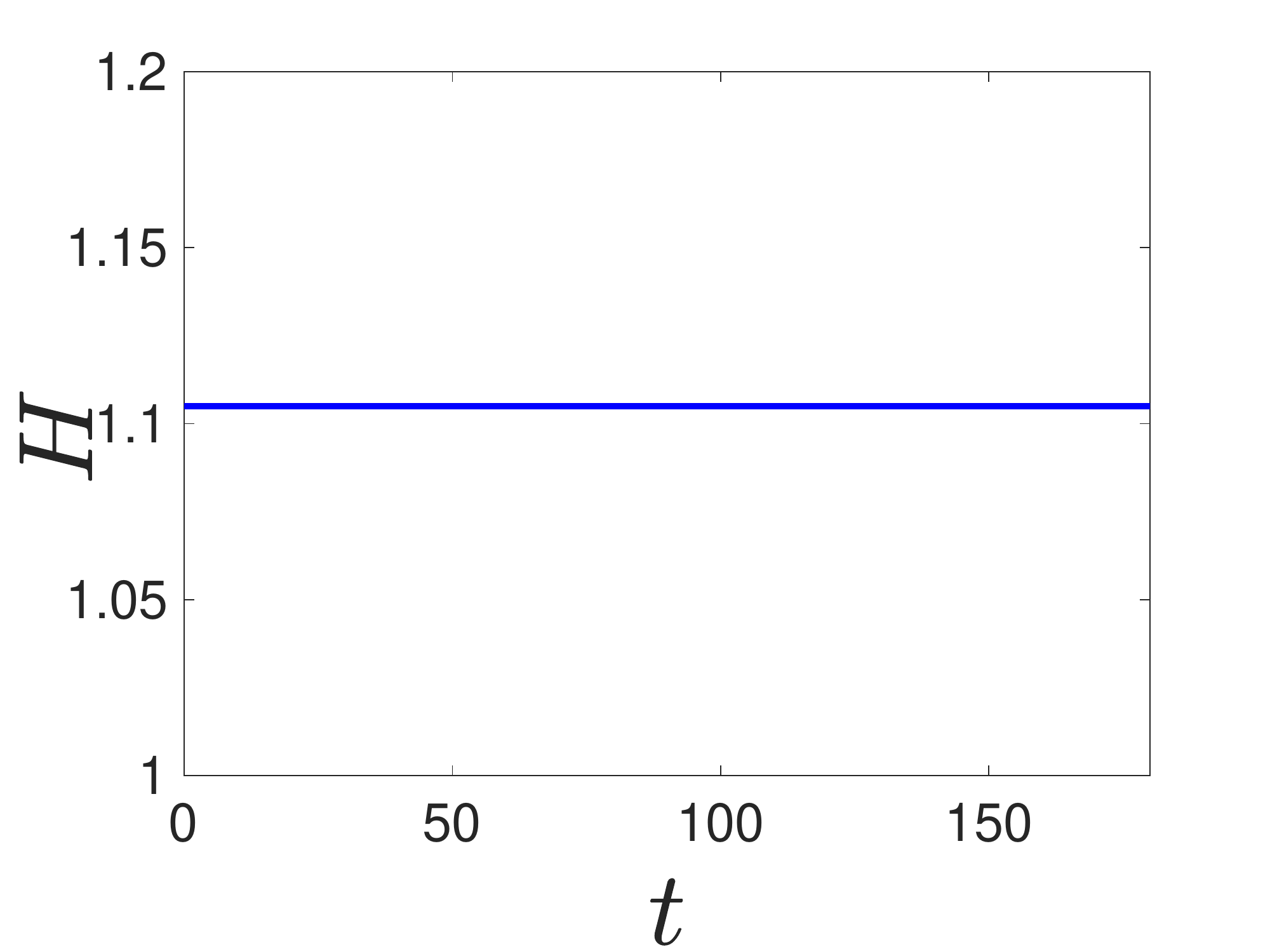}
\includegraphics[width=0.46\textwidth]{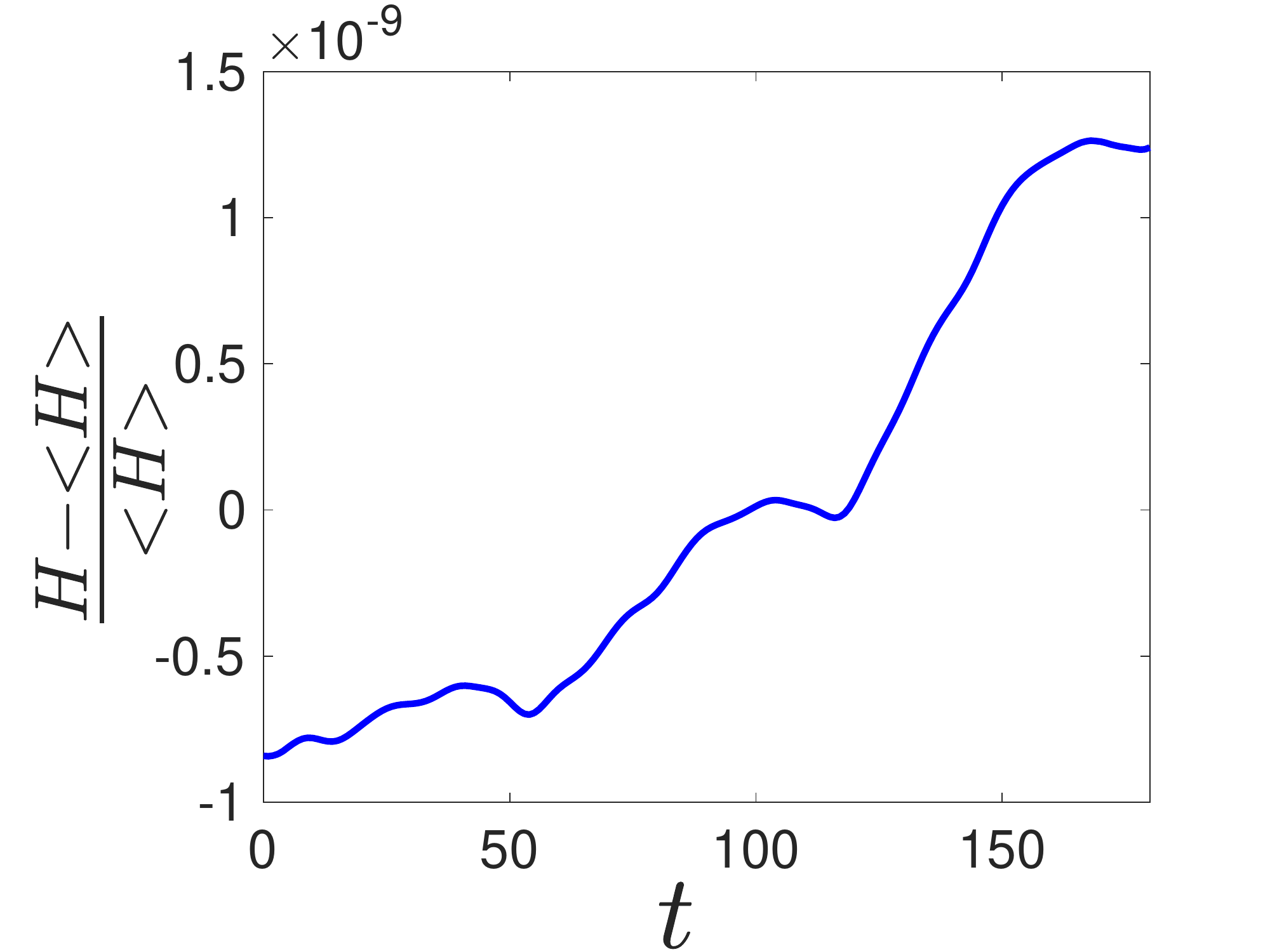}

\end{center}
\caption{An example of the evolution of the Hamiltonian
  and its deviations from its mean value for a single moving kink with 
  $c=0.3$.
  The left panel shows the total energy $H$ and the
  right panel shows the deviation around the mean $<H>$. }
\label{energy}
\end{figure}

It is known that the Eq. (\ref{beam}) has Hamiltonian structure, therefore it conserves an energy (Hamiltonian) functional given by
\begin{align}
 H = \mathcal{T}(u;t) + \mathcal{V}(u;t) = \int_{-\infty}^{+\infty} \left(\frac{1}{2}u_t^2 + \frac{1}{2}u_{xx}^2 + V(u)\right) \,\mathrm{d}x ,
\label{eq:H_par}
\end{align}
where the kinetic $\mathcal{T}$ and potential $\mathcal{V}$ energy contributions of the field, respectively, are
\begin{align*}
\mathcal{T}(u;t) &= \frac{1}{2} \int_{-\infty}^{+\infty} u_t^2 \,\mathrm{d}x,\\
\mathcal{V}(u;t) &= \int_{-\infty}^{+\infty} \left(\frac{1}{2} u_{xx}^2 + V(u)\right) \mathrm{d}x.
\end{align*}
Since $\mathrm{d}{H}/\mathrm{d}t=0$, ${H}$ is a given constant for a
chosen initial field configuration. In our simulations, the average
value of ${H}$ is of $\mathrm{O}(1)$, while the deviations from the
mean are (for the numerous examples we considered) no more than
$\mathrm{O}(10^{-9})$. In this way, we use energy conservation as a
partial
check of the validity of our numerical results. In Fig. \ref{energy},
we show a moving single kink with the speed $c=0.3$. The bottom left
and bottom right panels show, respectively, the total energy $H$ and the deviation
from the mean value $<H>$ (calculated over the time horizon
of our entire numerical computation).

\subsubsection{Conservation of Momentum}

\begin{figure}[H]
\begin{center}
\includegraphics[width=0.46\textwidth]{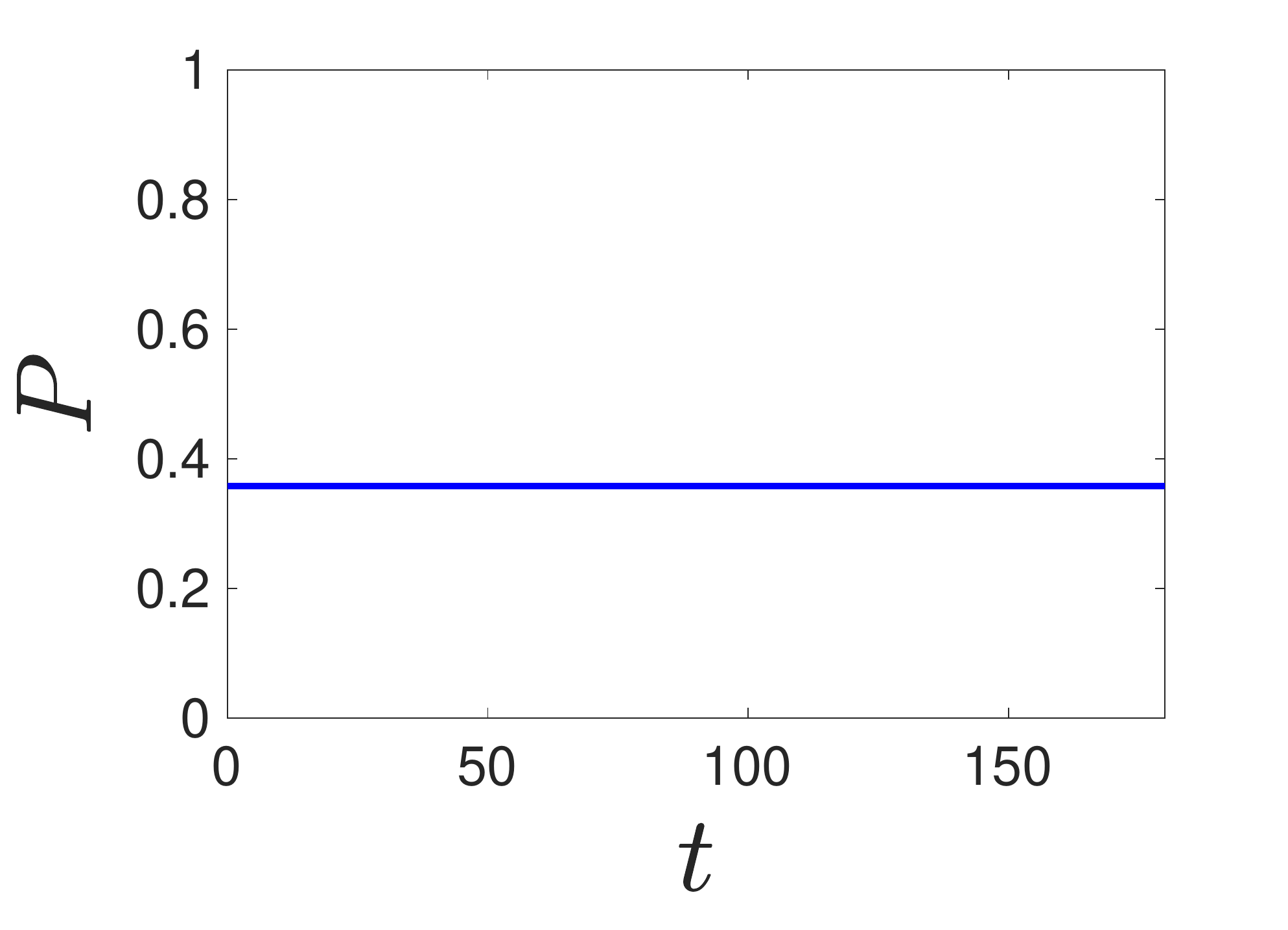}
\includegraphics[width=0.46\textwidth]{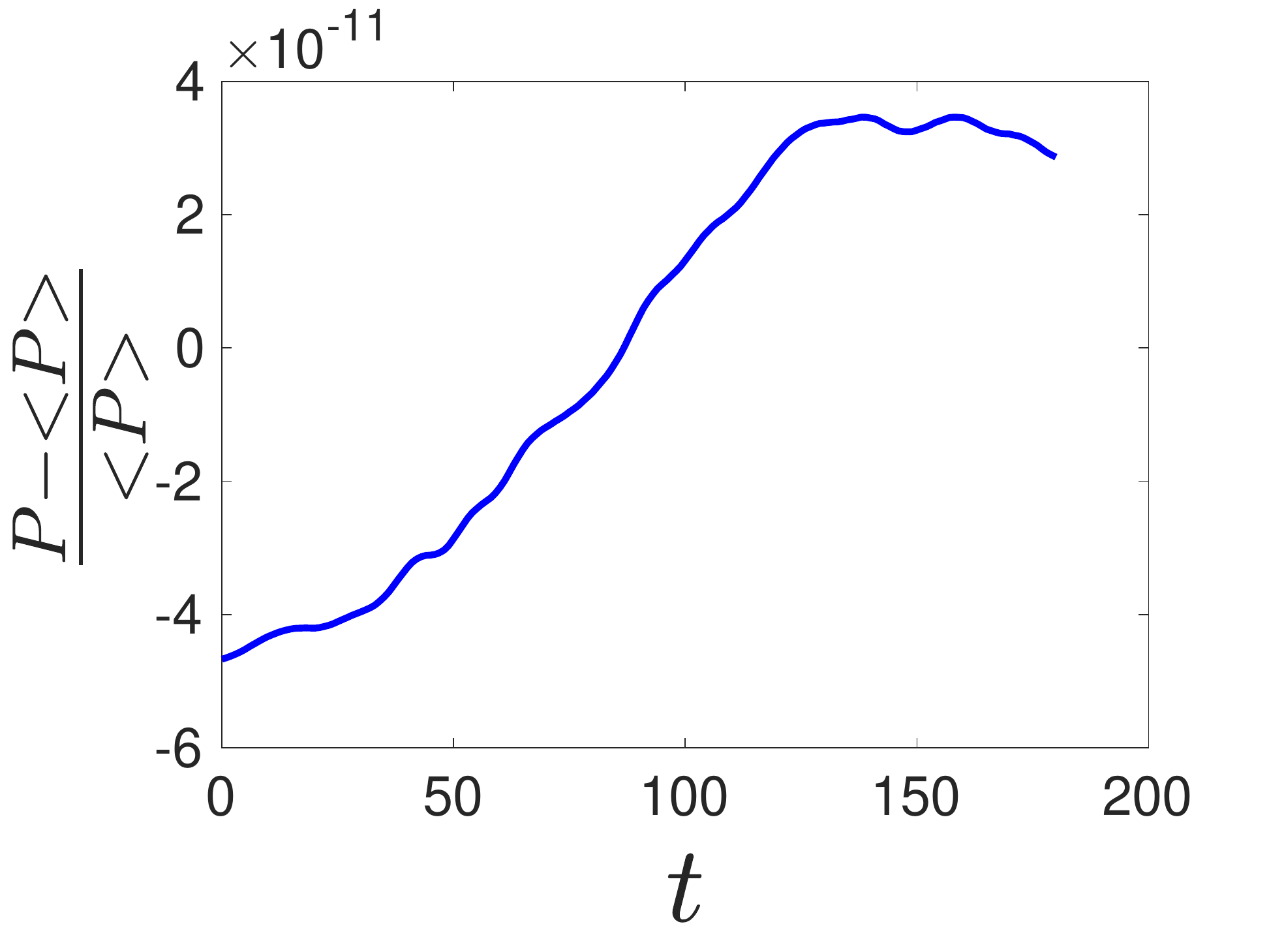}
\end{center}
\caption{Left panel shows the total momentum $P$ and the right panel
  shows the
  deviation aroundthe mean $<P>$ for a single moving kink with $c=0.3$. }
\label{momentum}
\end{figure}

Similarly to the energy, another important conservation law of the
B$\phi^4$ equation is that of the linear momentum (associated also
with
the invariance of the kink structures we discussed above with respect
to translations).
The momentum on the interval $(a,b)$ is defined as $\displaystyle P=-\int_a^b u_t
u_x \mathrm{d}x$. Differentiating $P$ with respect to time $t$, it is
straightforward
to infer that the relevant quantity is conserved.
In Fig. \ref{momentum},  we show the total momentum $P$ and the
deviation from the mean $<P>$ for a moving single kink with the speed
$c=0.3$. Once again the relevant quantity is of order unity, while the
deviations from its mean value are of O$(10^{-11})$ indicating the
accuracy of our numerical computations.

\section{Kink-Antikink Collisions}
Lastly, and most importantly for our study of the properties of the
B$\phi^4$ model, 
we now turn our attention to the topic of kink-antikink solutions.
Recall that such collisions have been the topic of intense scrutiny in
the
regular $\phi^4$
model~\cite{Campbell,Belova,Ann,goodman,goodman2,weigel,weigel2}.
Importantly, the recent work of~\cite{weigel,weigel2} and the summary
of~\cite{cuevas}
suggest that the relevant topic is far from complete. Hence, this is
naturally a theme of principal interest within the (fourth derivative)
model discussed
herein, namely the B$\phi^4$ equation.

For the separation half-distance we choose $x_0=20$ and let the kink
and antikink approach each other at various velocities
($v_{\mathrm{in}}$), and then record the average velocity at which
they separate after the interaction ($v_{\mathrm{out}}$). To generate
initial conditions,
we follow a technique that we developed in an earlier
work~\cite{christov}.
In particular, we use Matlab's \textit{lsqnonlin} to find $\varphi_{min,c}(x)$
which minimizes the quantity
$||\varphi^{(4)}+c^2\varphi''+V'(\varphi)||_2^2$ (square of the $\ell _2$-norm of the left side of Eq. (\ref{beam_traveling})) subject to the
additional constraints that the kink position remain at $x=-20$ and the
antikink at $x=20$. This is necessary because Eq. (\ref{beam_traveling}), which applies to a traveling wave solution to Eq. (\ref{beam}), may not have a solution when a kink and antikink are involved (for a single kink or antikink a solution is always possible). Thus a least-squares approximation is the best one can do. In this way, we ensure that the initial conditions produce the ``best'' possible approximation to a B$\phi^4$ kink and antikink traveling towards each other, each with speed $c$,
and consequently produces the minimal possible radiation as
a result of the coherent structure ``superposition''.

\begin{figure}[H]
\begin{center}
\includegraphics[width=0.48\textwidth]{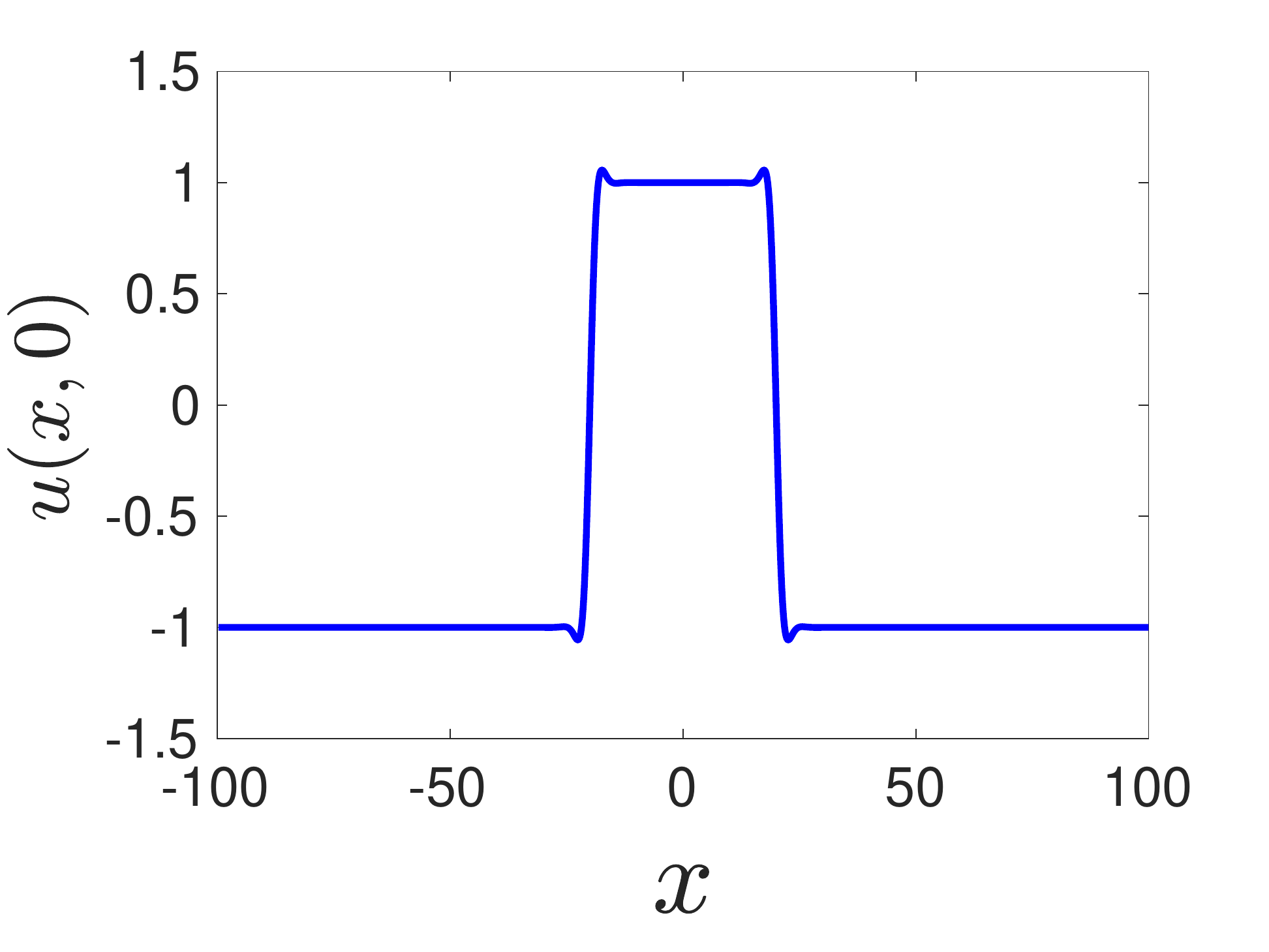}
\includegraphics[width=0.48\textwidth]{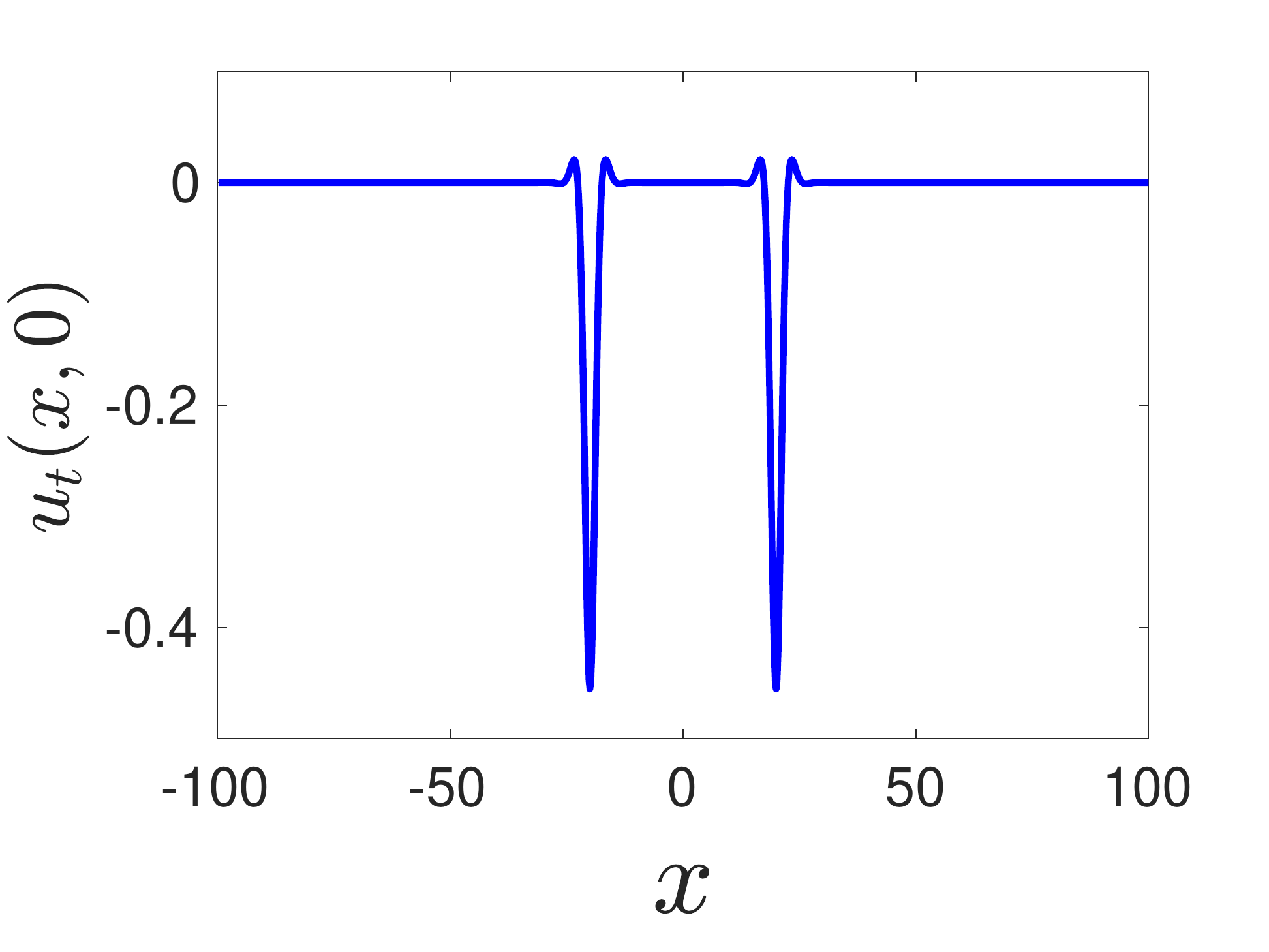}
\end{center}
\caption{Initial conditions for kink-antikink solution with $x_0=20$
  for $|v_{\mathrm{in}}|=0.55$. The left panel shows $u(x,0)$ and
  right one shows the
  plot for $u_{t}(x,0)$.}
\label{critical}
\end{figure}

As  initializer to \textit{lsqnonlin}, similar to \cite{christov}, we use 
\[u(x)=u_{c}(x+x_{0})+H(x)(u_{c}(x-x_{0})-u_{c}(x+x_{0}))\]
where $u_c(x)=\tanh(\frac{x}{\sqrt{1-c^2}})$ is the traveling wave
solution to Eq. (\ref{phi4}) at $t=0$. Here, $H(x)$ is the Heaviside
unit-step function. Then the initial conditions  that we
use for moving kink-antikink system are:
\begin{align*}&u(x,0)=\varphi_{min,c}(x) \\
&u_t(x,0)=-c \> \text{sign}(x) \varphi_{min,c}'(x)
\end{align*}
Note that without the ``sign'' function, the kink and antikink would move in the same direction. {See Figure \ref{critical} for a typical initial position $u(x,t=0)$ profile and initial velocity $u_t(x,t=0)$ profile}.

It is relevant to recall here the particularly complex phenomenology
of the regular $\phi^4$ model. There, sufficiently large velocities
($v_{\mathrm{in}}>0.2598$), the kink and antikink always inelastically
scatter,
while for sufficiently small ones ($v_{\mathrm{in}}<0.193$) they
always trap each other into a breathing, so-called bion, state.
In between, a remarkable wealth of fractal in nature multi-bounce
(2-bounce, at the edge of which there exist 3-bounce, at the end
of which 4-bounce, and so on) windows arise. In these, the coherent
structures, despite the (kinetic) energy loss they incur during the
first collision, they manage to escape each other's attraction via
a resonance mechanism involving the kink's internal mode after
multiple (respectively, 2-, 3-, 4-) bounces.

\begin{figure}[H]
\begin{center}
\includegraphics[width=0.48\textwidth]{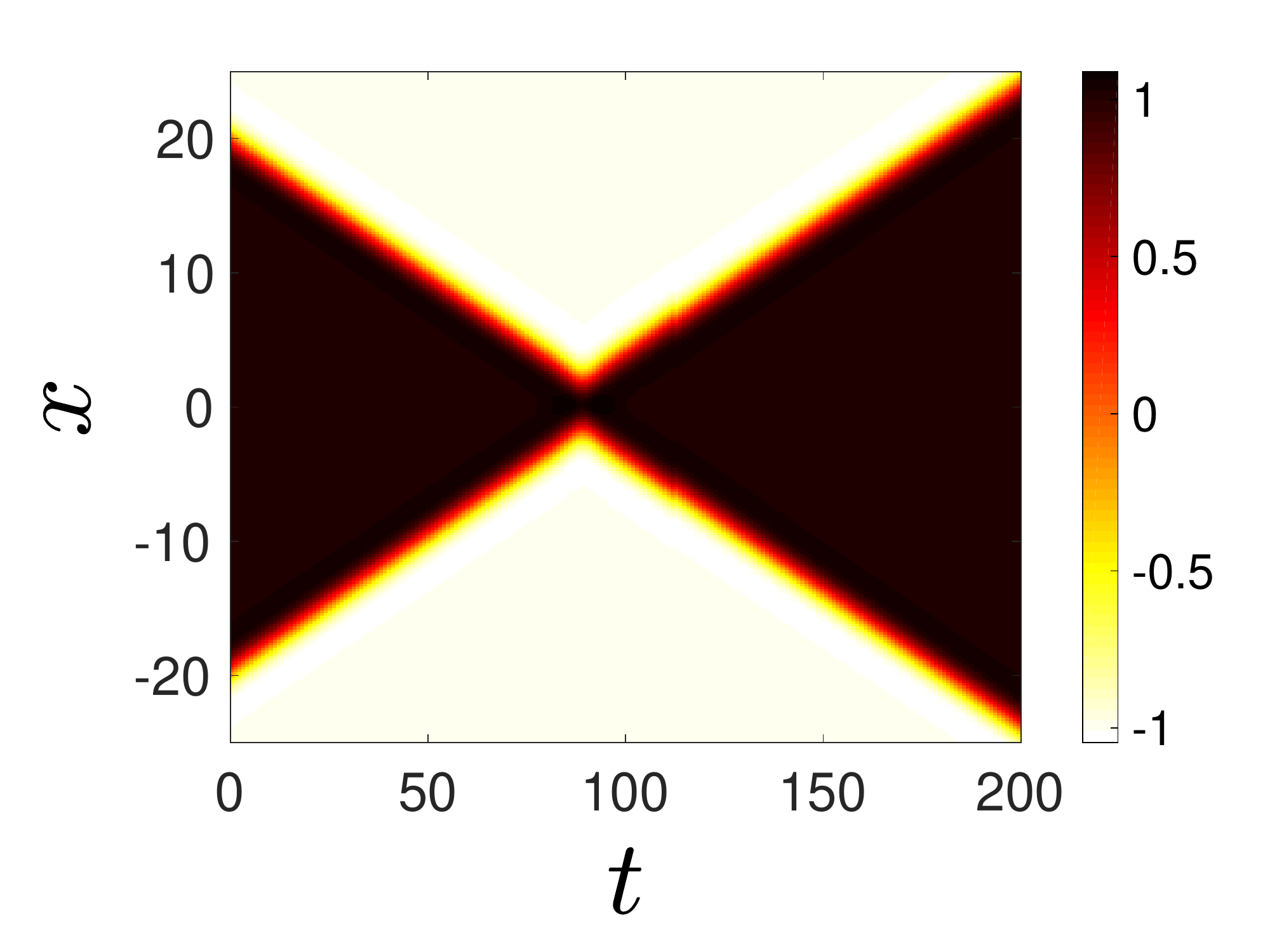}
\includegraphics[width=0.48\textwidth]{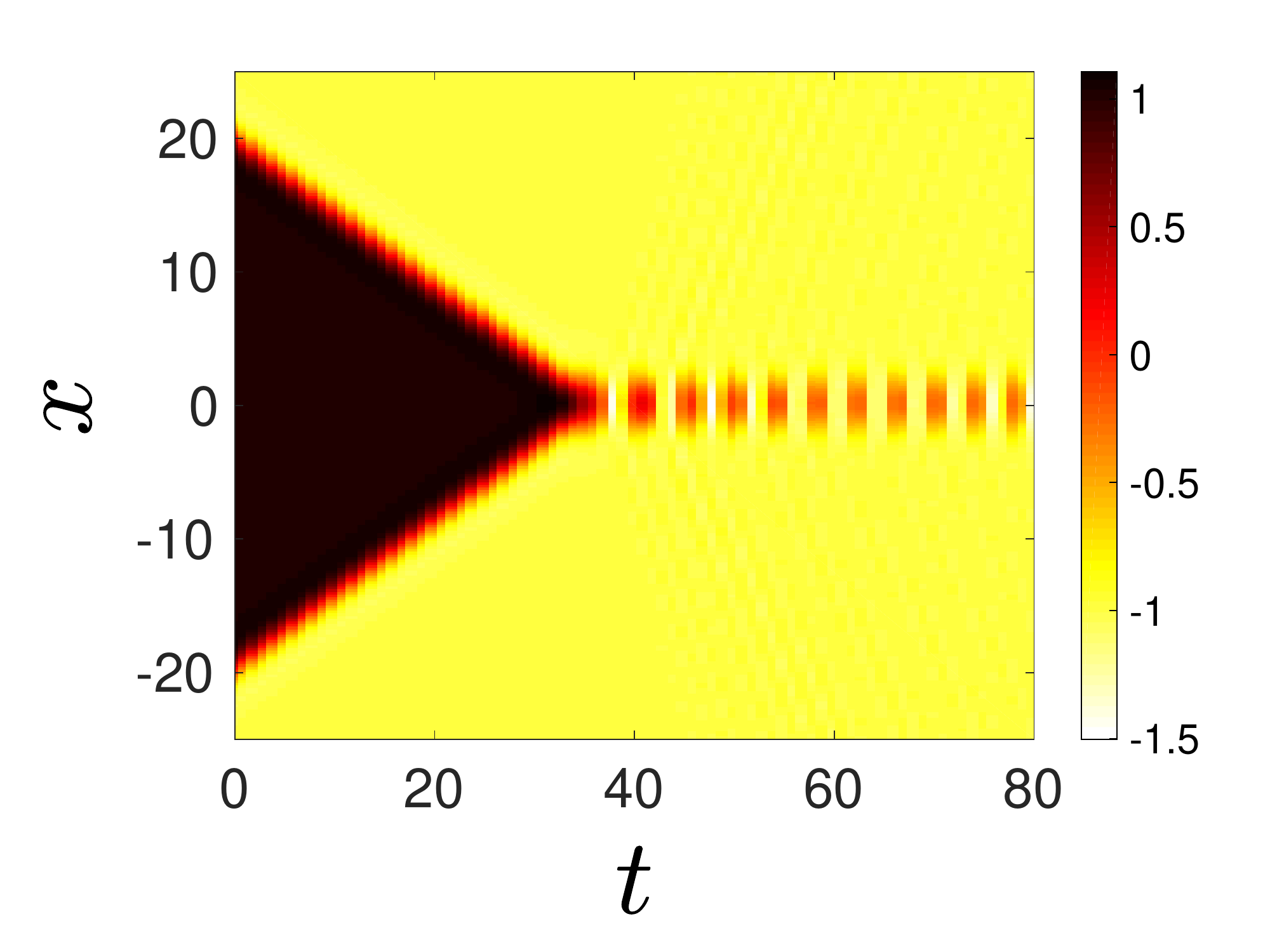}
\includegraphics[width=0.48\textwidth]{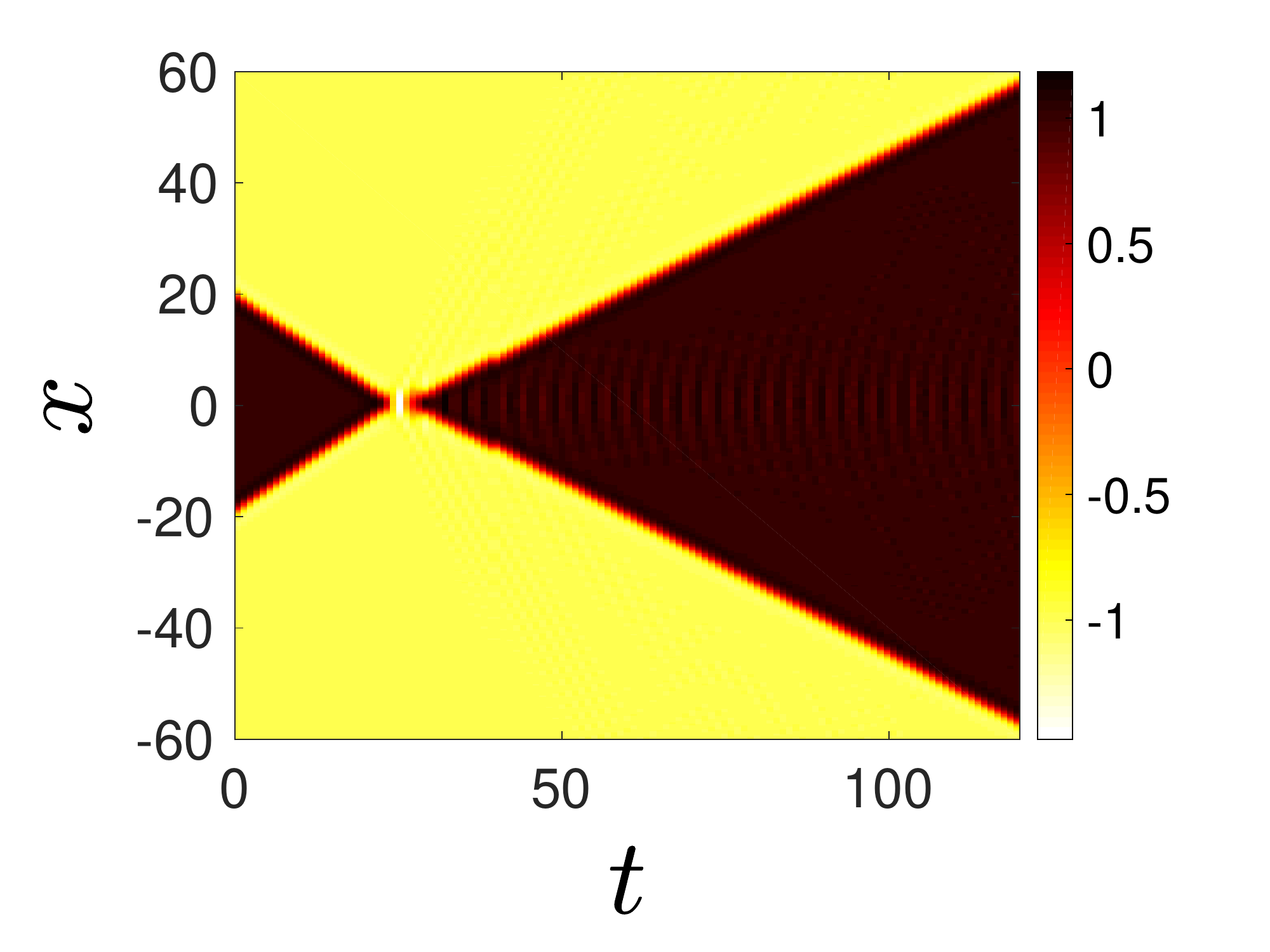}
\end{center}
\caption{The top left panel shows the repelling of the kink-antikink state
  ($|v_{\mathrm{in}}|=0.2$). The top right panel depicts
  the case when the kink and antikink collide infinitely many times
  ($|v_{\mathrm{in}}|=0.55$). The bottom panel shows an example of the case where they
  collide once and
  then escape from each other forever ($|v_{\mathrm{in}}|=0.8$).}
\label{threecases}
\end{figure}

The collision picture in the B$\phi^4$ model turns out to be
dramatically
different and while in some ways it is quite simpler, in others it
turns
out to also be rather complex. More specifically,
for most initial velocities used we end up with three cases. In the
first case, where $|v_{\mathrm{in}}| \in (0.001, 0.5108)$ (the no
bounce window), the kink and antikink move towards each other, but
after some certain time they stop and move away from each other. In
the second case where $|v_{\mathrm{in}}| \in (0.5109,0.5895)$ (the
infinitely many bounce window) the kink and antikink move towards each
other and collide, but they
do not have enough kinetic energy to escape from each other. They end
up with infinitely many collisions, i.e., trapping each other. In the third case, where $|v_{\mathrm{in}}| \in (0.5896, 1)$ (the one bounce window), the kink and antikink collide only once and they escape from each other forever as seen in Figure \ref{threecases}.

\begin{figure}[H]
\begin{center}
\includegraphics[width=0.48\textwidth]{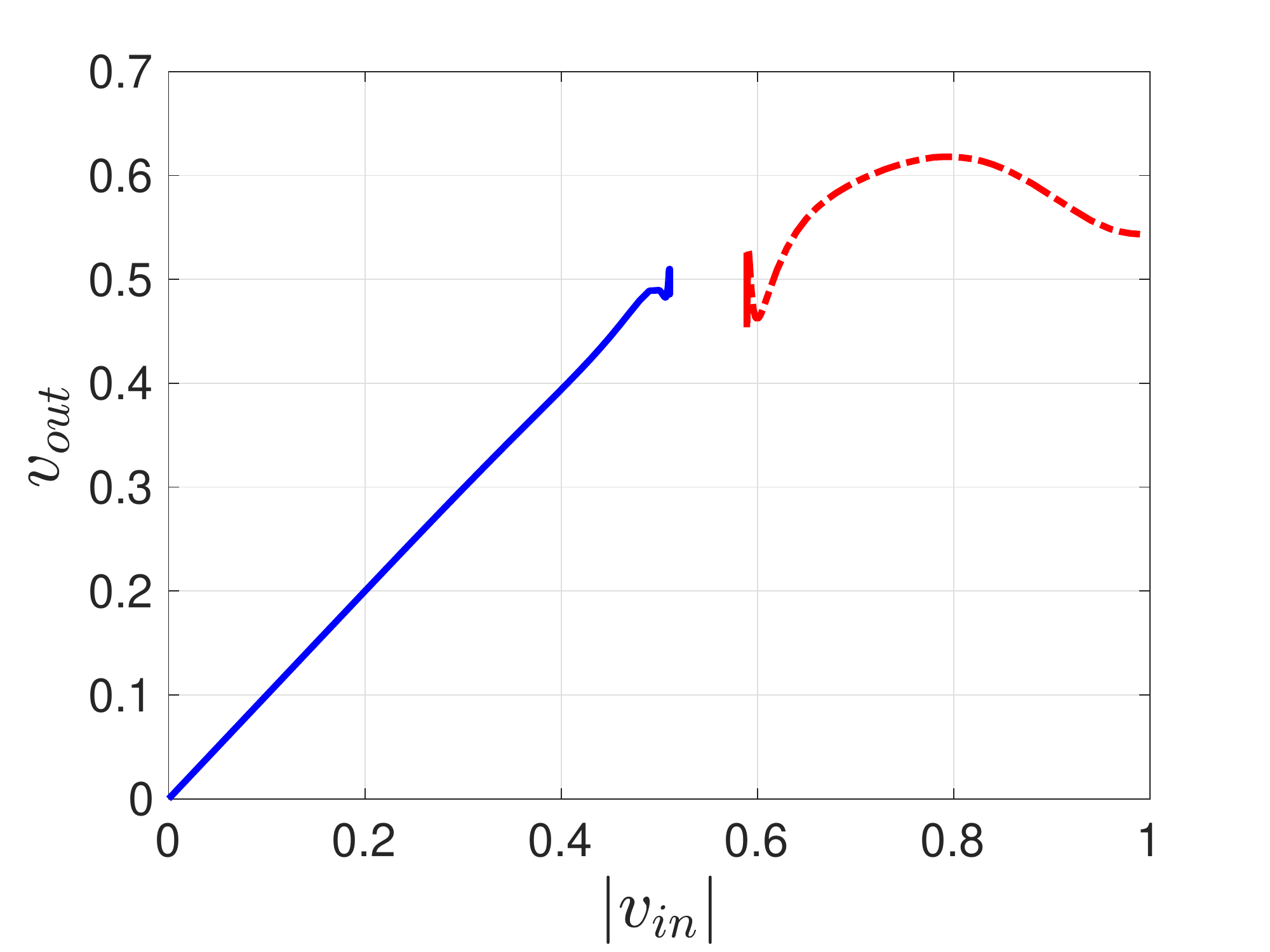}
\end{center}
\caption{The relation between  $v_{\mathrm{out}}$ vs $|v_{\mathrm{in}}|$. The blue solid line corresponds to the first case, where the kink and antikink repel each other. The red dashed curve corresponds to the third case, where the kink and antikink collide only once and then escape from each other.}
\label{vin_vout}
\end{figure}

It is clear from the nature of the interaction of the top left panel
of Fig.~\ref{threecases} that the kink and antikink
  effectively
``repel'' each other {when they get sufficiently close}.
That is to say if they do not possess sufficiently large speed, they
will not be able to overcome the energetic barrier that precludes
them from colliding. 
In Fig. \ref{vin_vout}, we present the relation between
$|v_{\mathrm{in}}|$ and $v_{\mathrm{out}}$. We observe that for small
values of $|v_{\mathrm{in}}|$, there is a linear relationship with
$v_{\mathrm{out}}$,
such that to a very good approximation
$v_{\mathrm{out}}$=$|v_{\mathrm{in}}|$ . We do not see a linear
relation for larger values of $v_{\mathrm{in}}$. This suggests that
small
kinetic energies (smaller than the one of the energetic barrier
precluding
the kink-antikink collision) will lead to direct reflection with
minimal
conversion to a different form of energy. 
On the other hand, if the waves are incoming with sufficiently large speed,
they
will collide and separate after a single bounce (bottom panel of
Fig.~\ref{threecases}). However, in that case, as shown in
Fig.~\ref{vin_vout},
the outgoing speed will be significantly smaller than the incoming one
signaling
the conversion of the kinetic energy into internal energy and also
importantly small amplitude dispersive radiation wavepackets. 

The most interesting case naturally lies between the two above limits.
Here the initial kinetic energy of the waves is
higher than the (repulsive) barrier, thus the structures will reach each other
and
collide. 
Our detailed numerical computations in the vicinity of the boundary
of such a collision have revealed a surprising feature. This
occurs near the boundaries of the infinitely-many bounce
window. Letting $v_L$ represent the left boundary of the
infinitely-many bounce window, and $v_R$ the right boundary of the
same window, we see that there appear to be oscillations in the
$v_{\mathrm{out}}$ versus $|v_{\mathrm{in}}|$ curve as
$|v_{\mathrm{in}}|$ approaches $v_L$ from the left and as
$|v_{\mathrm{in}}|$ approaches  $v_R$ from the right. Closer
inspection of these regions show that this is indeed the case.

\begin{figure}[H]
\begin{center}
\includegraphics[width=0.46\textwidth]{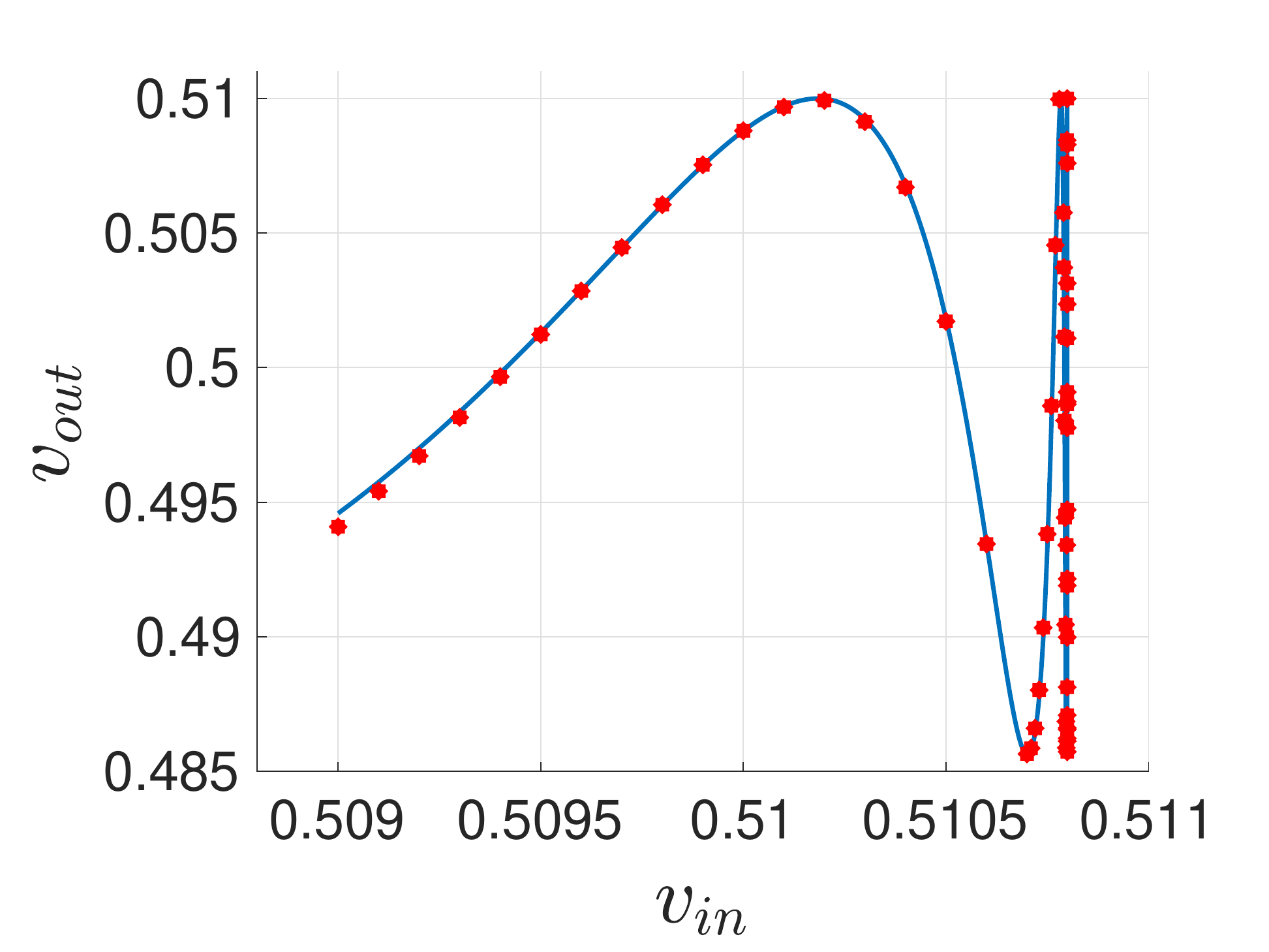}
\includegraphics[width=0.46\textwidth]{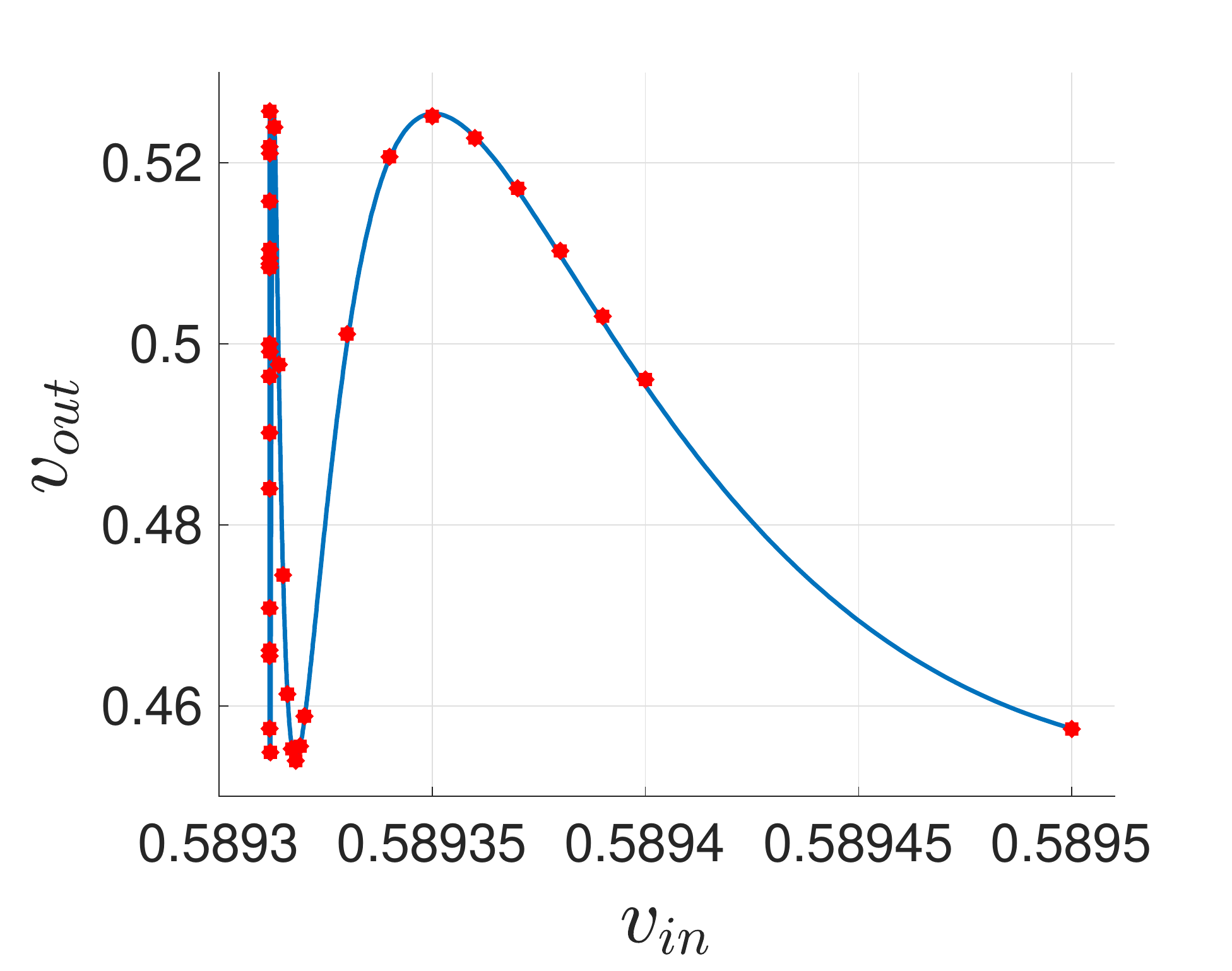}\\
\includegraphics[width=0.46\textwidth]{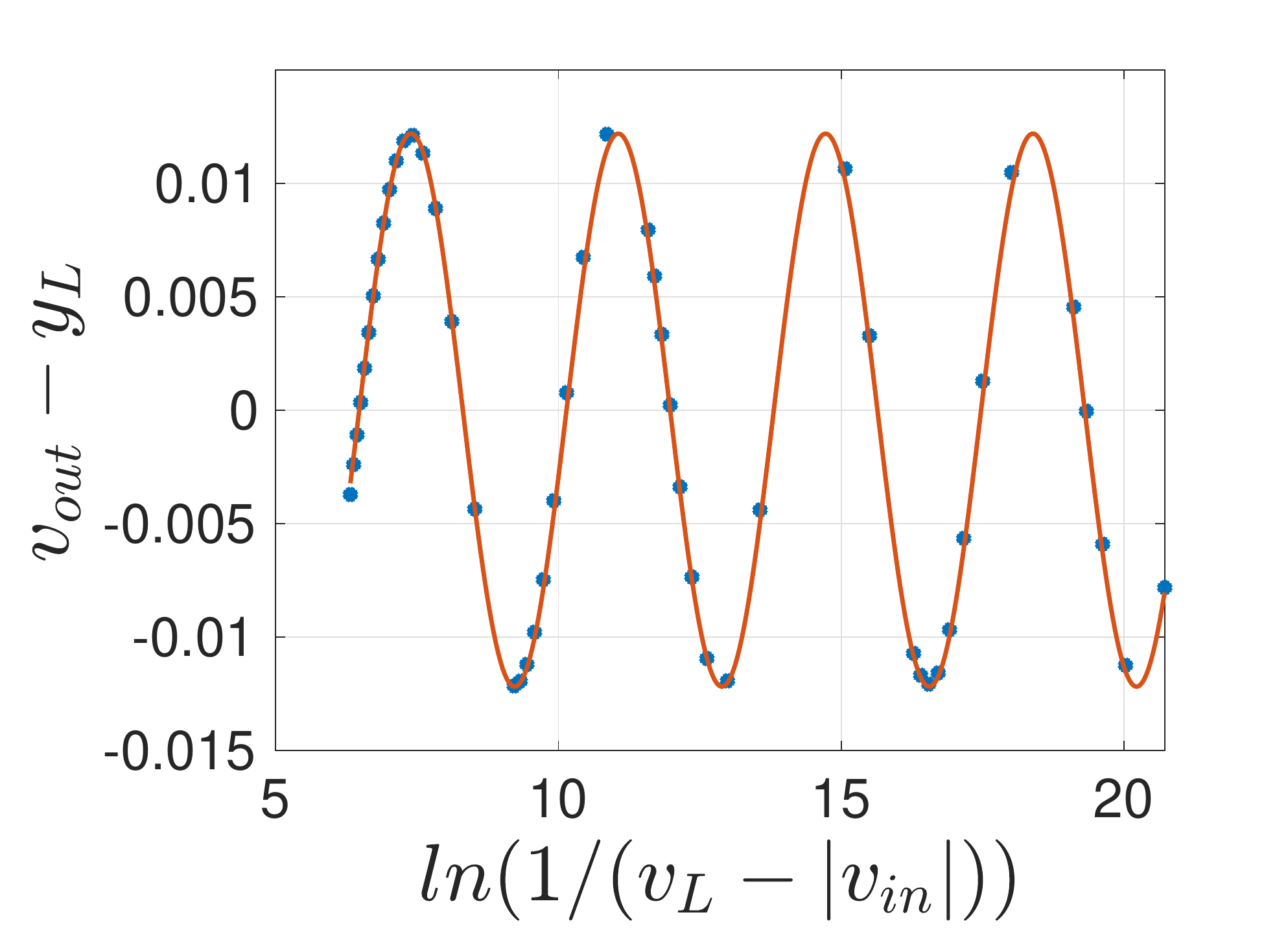}
\includegraphics[width=0.46\textwidth]{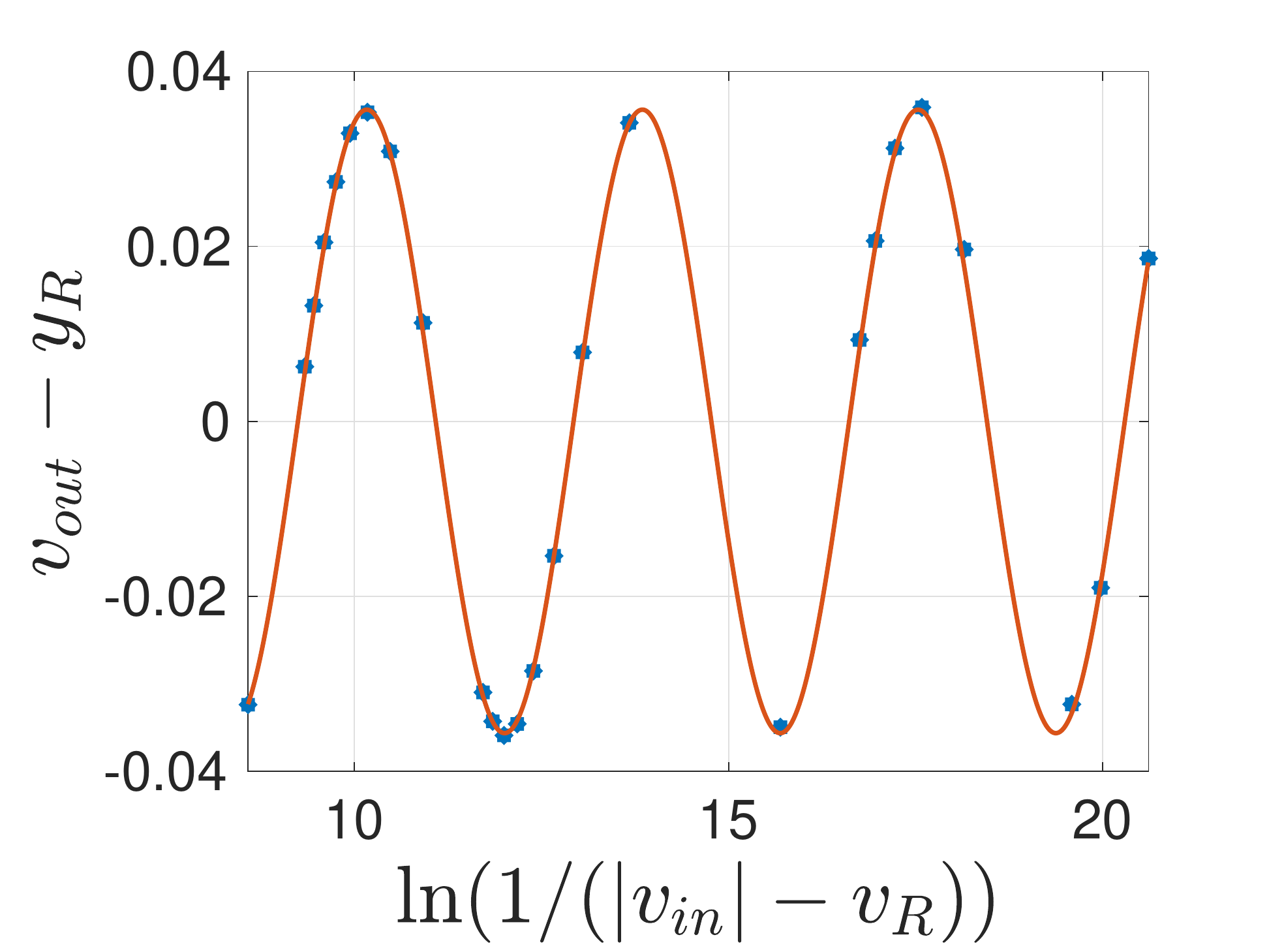}
\end{center}
\caption{Oscillations near the edges of the infinitely-many bounce
  window.
  Upper left: original data and fitted model, left critical value. Upper right: original data and fitted model, right critical value. Lower left:  transformed data and fitted model, left critical value. Lower right: transformed data and fitted model, right critical value.}
\label{oscillationsCrit}
\end{figure}

In Figure \ref{oscillationsCrit} we show close-up views of these two
regions (top two panels). In both cases we observe oscillations that
get more rapid as the critical point ($v_L$ or $v_R$) is
approached. Upon a systematic data exploration, it was found that the
data follows
a pattern similar to that of $\sin(\log(1/|x|))$ as $x \rightarrow 0$. Thus it appears that no limit for $v_{\mathrm{out}}$ exists as $|v_{\mathrm{in}}|$ approaches $v_L$ from the left or $v_R$ from the right. This is in stark contrast to the corresponding $\phi^4$ model given in Eq. (\ref{phi4}). For that model, we know that $v_{\mathrm{out}}$ always goes to zero at the boundaries of any n-bounce window. 
The $v_{\mathrm{out}}$ versus $|v_{\mathrm{in}}|$ data near each
critical point ($v_R$ and $v_L$) was first translated to the origin
(i.e., $v_L$ or $v_R$ was respectively subtracted),
then $\log(1/x)$, ($x=$ the translated $v_{\mathrm{in}}$) was plotted
against the
translated $v_{\mathrm{out}}$ data. The results are in the bottom two
panels of Figure \ref{oscillationsCrit}. Since the pattern of the data
appears sinusoidal, a numerical fit to a sine function of the form
$a\sin(bx+c)$ was performed (with $x$ representing the transformed
$v_{\mathrm{in}}$ data). The results appear in the four panels of
Figure \ref{oscillationsCrit}. In the bottom two panels the
transformed data and fitted functions appear, and in the top two
panels the original data and the model for the data (derived from the
fitted functions in the bottom panels). In all cases the models fit
the data quite well ($R^2=0.99$ or higher). This suggests a very
delicate oscillatory regime of outgoing velocities both on the side
of a of $v_L$ and on that of $v_R$.

\begin{figure}[H]
\begin{center}
\includegraphics[width=0.48\textwidth]{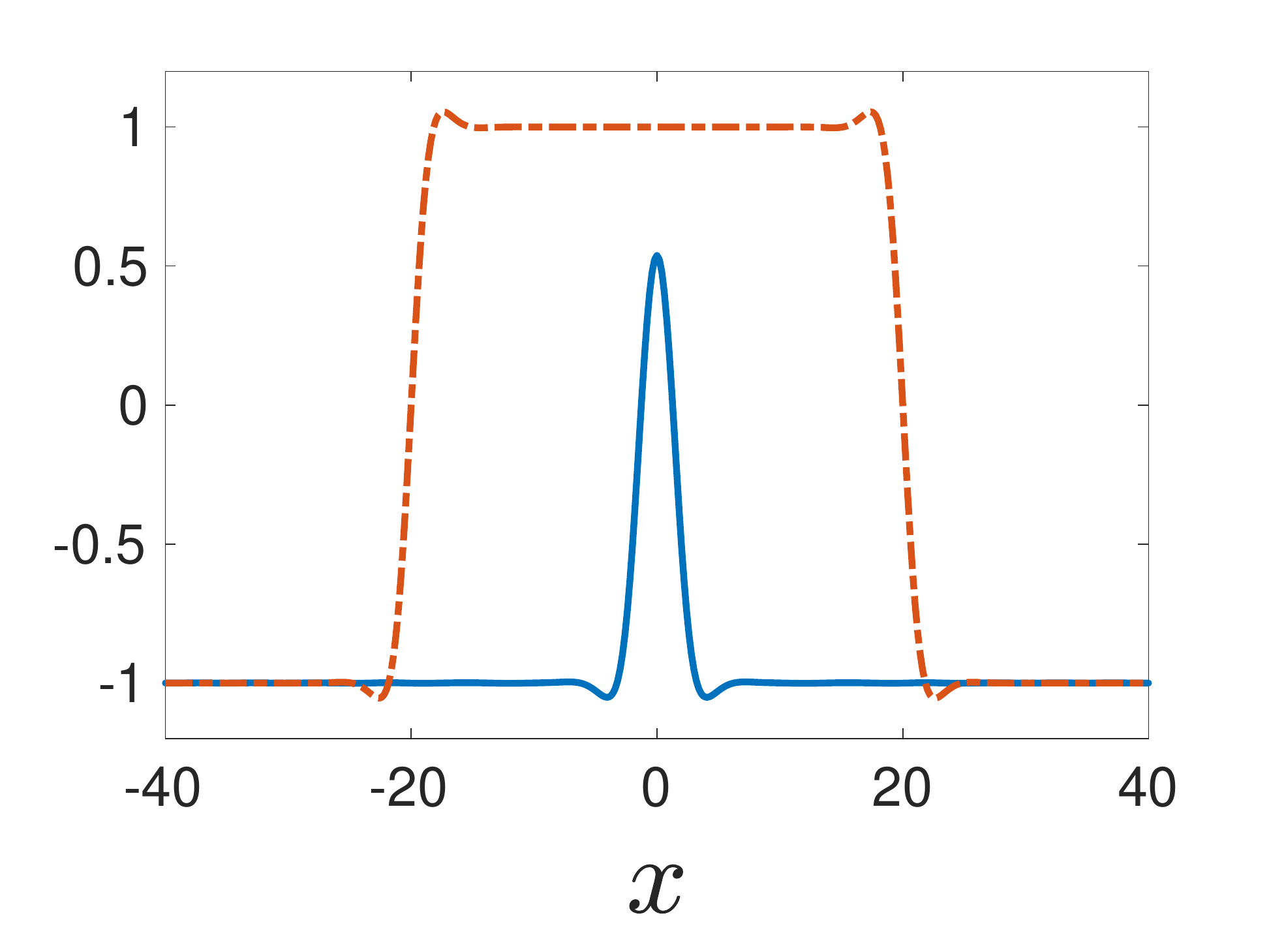}
\includegraphics[width=0.48\textwidth]{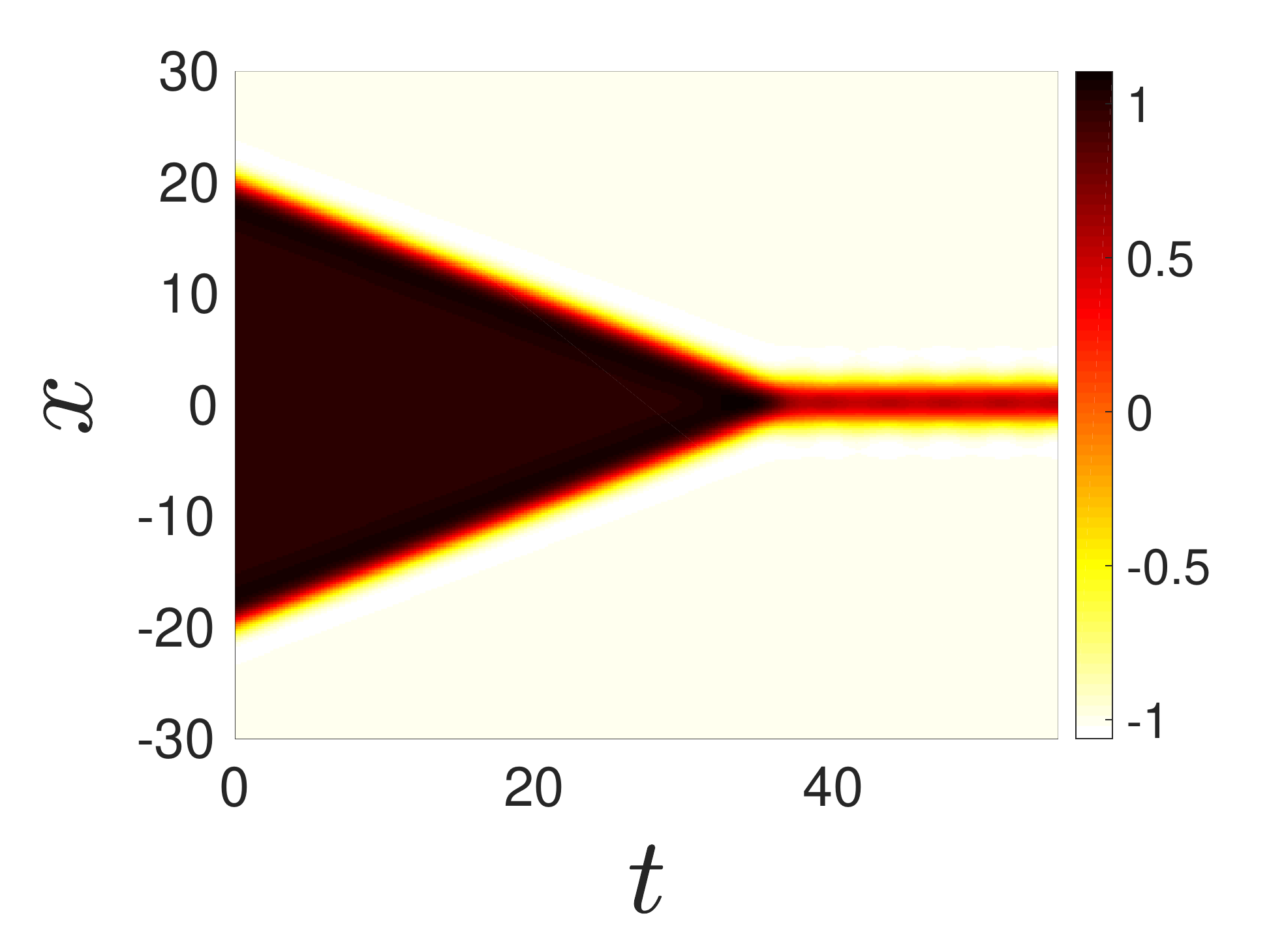}
\includegraphics[width=0.48\textwidth]{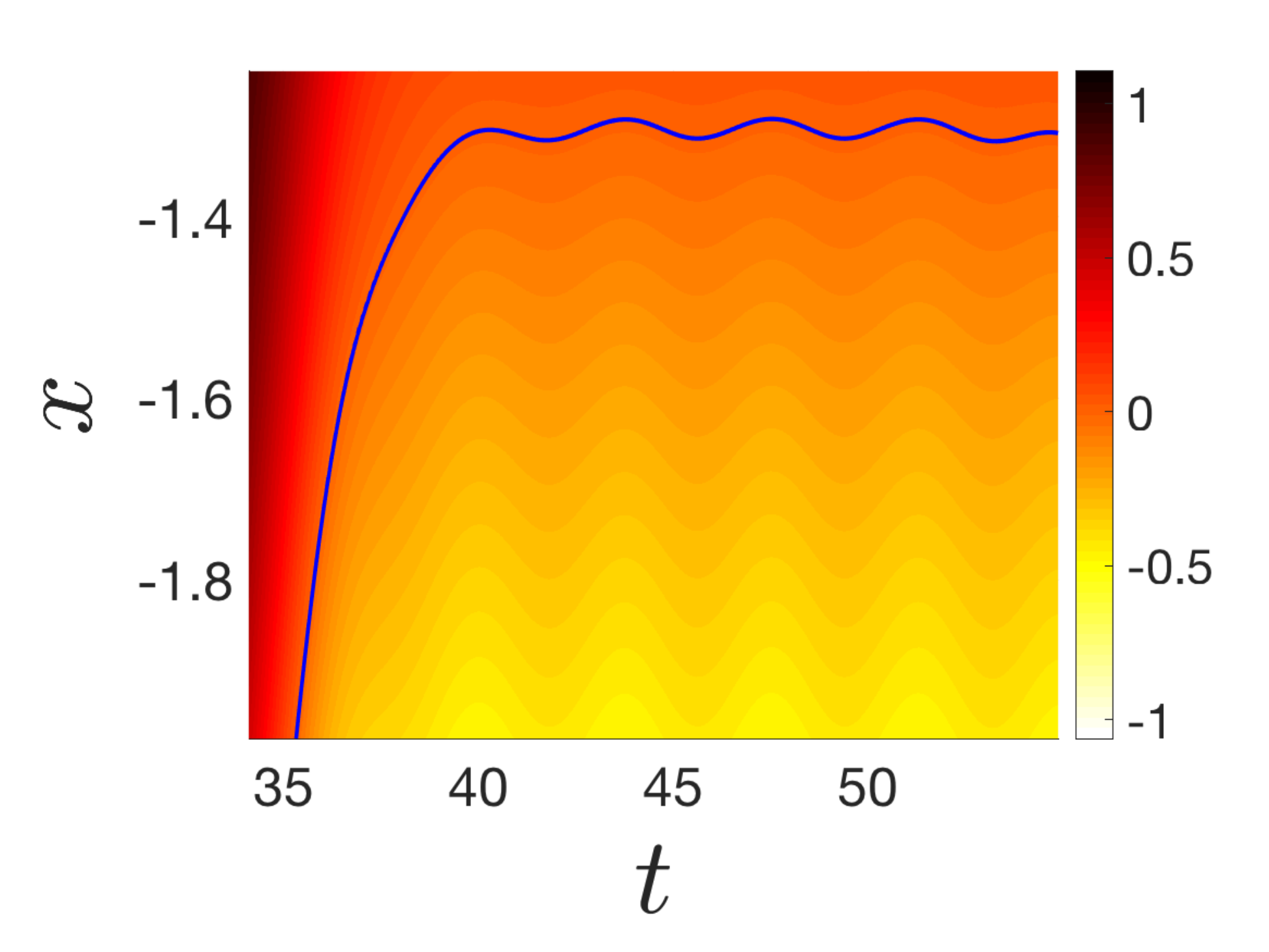}
\end{center}
\caption{Kink-antikink solution with $x_0=20$ for
  $|v_{\mathrm{in}}|=0.510799$ on the time interval $[0,55]$. On the upper
  left panel, the red dashed line represents the kink-antikink pair
  when $t=0$ and the blue solid line represents the kink-antikink pair
  when $t=55$. The upper right panel shows the contour plot of this
  kink-antikink state on the time interval $[0,55]$. The lower panel is a zoom of the upper right panel with the position of the kink suprimposed (blue solid line).}
\label{critical2}
\end{figure}

We end this section with a bit of a speculation about the source of
these oscillations.
While the regimes of individual behaviors of the B$\phi^4$ model
are
far fewer and more well defined than in the second derivative $\phi^4$
analogue, these oscillations are a source of unexpected complexity.
In Figure \ref{critical2} we show the results of using the initial
conditions $|v_{\mathrm{in}}|=0.510799$ and $x_0=20$ (the dashed red
curve shows
the initial position in the first panel) that result in a kink-antikink pair approaching what appears to be a steady state (blue curve in first panel). The second panel is a contour plot showing that this apparent steady state develops at approximately $t=35$ and persists to at least $t=55$. Near the other critical $|v_{\mathrm{in}}|$ value of about 0.5896, we also observe that the kink-antikink solitions appear
to reach a steady-state for some time (in a similar manner, hence
omitted here). In fact, the combined kink-antikink state is oscillating
slightly about the steady state shown in the left panel which can be
seen in an enlargement of the contour plot in the region $35\le t \le 55$, shown in the lower panel in Figure \ref{critical2}. Thus for very small
changes in $v_{\mathrm{in}}$ near the critical values (but not entering the range between the two critical values), the oscillating solitons will separate
at different points in their oscillatory cycles, resulting in the different (oscillating)
outgoing velocities $v_{\mathrm{out}}$.

Finally we note that with very small perturbations in
$v_{\mathrm{in}}$ which do enter the region between the critical values, we observe that after the kink-antikink pair undergoes small oscillations about a steady state for a while, they get
stuck with infinitely many collisions (bion state). This suggests
that
in addition to the potential barrier discussed above, there exists
also a bound state in the form of a potential well that can trap the multi-kink dynamics.
The oscillatory structure of the outgoing velocities outside the region between the critical values is indicative of
the possibility that multiple such equilibrium states (saddles and
centers)
may exist.
Exploring the structure and stability of these steady states (as
dictated by the oscillatory nature of the kink tails) will be a subject of
future work. 

\subsection{Collective Coordinates Method (ODE)}

One of the prototypical methods that have been used to attempt to
understand
the dynamics of the $\phi^4$ model is the collective coordinate (CC)
method.
Here, the evolution of the kink and antikink is represented by a
suitable superposition ansatz featuring a finite
number of time-dependent collective variables (such as the center and width of the
kinks or the amplitude of their internal mode) and the evolution of
the
ODEs for these variables is developed (typically) based on the
underlying
Lagrangian of the PDE model. In this setting the original analysis
of~\cite{Sugiyama}
was used later, e.g., by~\cite{Ann} and further in a quantitative
fashion
in~\cite{goodman,goodman2}. However, recently, the work
of~\cite{weigel,weigel2} revealed some inconsistencies in the original
ODE derivation of~\cite{Sugiyama} leading to the need for
reconsideration
of the entire CC framework for the $\phi^4$ model.

Here, our scope is more modest, as we will only illustrate how to
consider the setting with a single collective coordinate, namely the
center of the kink and antikink. As we will discuss further below,
while partially useful in the B$\phi^4$ model, this approach has
nontrivial limitations that are worthwhile to further explore and amend in
future studies.
Our aim  is to reduce the full PDE with infinitely many degrees of
freedom to a simple model with only one degree of freedom and explore
the potential successes and the nontrivial limitations of such an approximation. 

Assuming that we characterize the
kink-antikink motion by utilizing the ansatz
\begin{equation}
\label{ansatz}
  u(x,t)=\varphi_0(x+X(t))-\varphi_0(x-X(t))-1\end{equation}
where $\varphi_0(x+X(t))$ is the steady state kink solution of
Eq.~(\ref{beam}) whose center is located at $x=-X(t)$ and
$-\varphi_0(x-X(t))$
is the steady state antikink solution whose center is located at
$x=X(t)$. Note that the steady state solution centered at
$X(t)=0$, i.e. $\varphi_0(x)$ is shown in Fig. \ref{steady}.  
Our aim is to study the behavior of $X(t)$ with the initial conditions
$X(0)=x_0$ and $X'(0)=v_{\mathrm{in}}$ where $x_0$ is the distance
from the origin, and $v_\mathrm{in}$ is the initial speed of the
kink. Using  the Lagrangian of the PDE model in the form:
\begin{equation}\label{lagr}
\begin{aligned}
\mathcal{L}(u;t) &= \mathcal {T}(u;t) - \mathcal {V}(u;t)\\
&= \int_{-\infty}^\infty  \left(\frac{1}{2}u_t^2 - \frac{1}{2}u_{xx}^2 - V(u)\right) \, dx
\end{aligned}
 \end{equation}
we substitute the ansatz of Eq.~(\ref{ansatz}) to obtain:

\begin{equation}
\begin{aligned}
\mathcal{L}(u;t) &= \int_{-\infty}^\infty  \left(\frac{1}{2}u_t^2 - \frac{1}{2}u_{xx}^2 - V(u)\right) \, dx\\
&=b_0(X)\dot X^2 - b_1(X).
\end{aligned}
\end{equation}
Here
\begin{equation}
\begin{aligned}
b_0(X) = \frac{1}{2} \int_{-\infty}^\infty  &(\varphi_0'(x+X(t))+\varphi_0'(x-X(t)))^2 \, dx\\
b_1(X) = \frac{1}{2} \int_{-\infty}^\infty &(\varphi_0''(x+X(t))-\varphi_0''(x-X(t)))^2 \, dx \\
+ \int_{-\infty}^\infty  &V(\varphi_0(x+X(t))-\varphi_0(x-X(t))-1) \, dx.\\
\end{aligned}
\end{equation}
By applying the Euler-Lagrange prescription
\begin{equation}\label{euler-lagrange1}
\begin{aligned}
\frac{\partial \mathcal L}{\partial X} - \frac{d}{dt}\bigg(\frac{\partial \mathcal L}{\partial \dot X}\bigg) = 0,
\end{aligned}
\end{equation}
we obtain the dynamical evolution:
\begin{equation}\label{eq:EL}
\begin{aligned}
&\dot X = Y \\
&\dot Y = -\frac{1}{2}\frac{b'_0(X)}{b_0(X)}Y^2 - \frac{1}{2}\frac{b_1'(X)}{b_0(X)}.
\end{aligned}
\end{equation}

\begin{figure}[H]
\begin{center}
\includegraphics[width=0.46\textwidth]{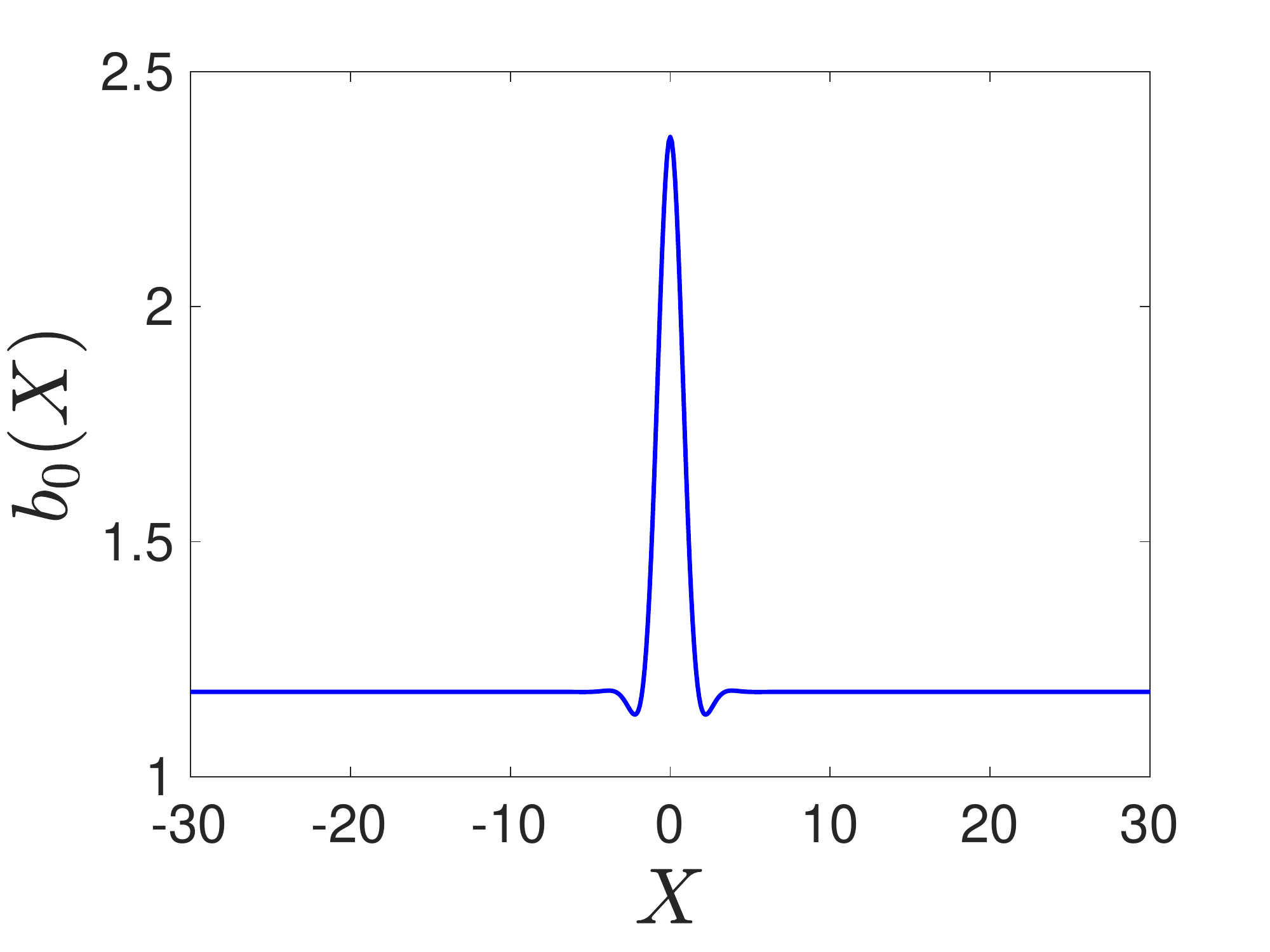}
\includegraphics[width=0.46\textwidth]{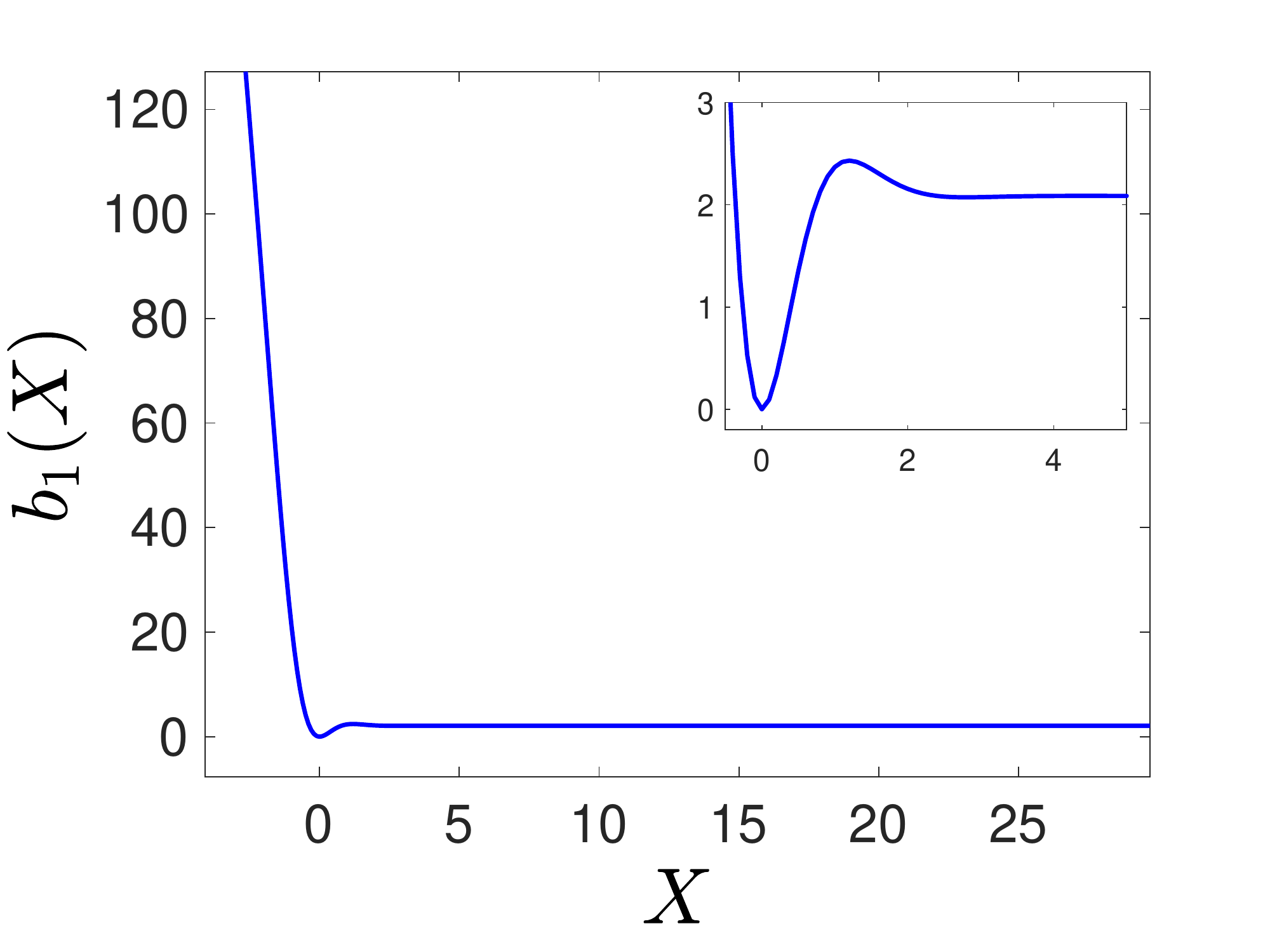}
\end{center}
\caption{The figure shows the plots of the coefficient functions $b_0(X)$ (left) and $b_1(X)$ (right).}
\label{coeff_b0_b1}
\end{figure}

We solve these equations numerically by using the initial conditions $X(0)=x_0$ and $Y(0)=v_{\mathrm{in}}$. We numerically compute the integrals on the interval $[-200,200]$. We use MATLAB's built-in fourth-order Runge--Kutta variable-step size solver {\tt ode45} with built-in error control. In Fig. \ref{coeff_b0_b1}, we show the coefficient functions $b_0(X)$ and $b_1(X)$.

\subsubsection{Results}

The CC method gives a very good match with the PDE results when
$v_{\mathrm{in}}$ is small, that is when $v_{\mathrm{in}} \in
(0,0.25)$.  We observe a difference when we start increasing
$v_{\mathrm{in}}$. This difference gets bigger as $v_{\mathrm{in}}$
gets closer to $0.51$. The relevant deviation becomes maximal when there is
an infinite bounce window in the B$\phi^4$ PDE simulations.
It is important to appreciate that
the CC method cannot  capture those bounces.
Bearing a single degree of freedom (dof) and given the conservation of
energy, the CC method can at best capture a pair of kinks that
interact
and become outgoing ones with the same speed as they were
incoming. Hence, beyond this threshold where the phenomenology deviates
from this symmetric scenario, the reduction of the PDE to the 1-dof
manifold
is one that is too restrictive to capture the relevant dynamics.
For bigger values of $v_{\mathrm{in}}$, we only see a good match until
the kink and antikink collide. After the collision, in the CC method,
as described above, the kinks separate from each other with a speed
practically equal  to
$v_{\mathrm{in}}$ whereas in the PDE  the kinks separate from each
other with a
speed that is smaller than $v_{\mathrm{in}}$. This inelasticity of the
collision
is due to the additional dof's of the B$\phi^4$ field theory which are
naturally not captured in this reduced CC formulation.

\begin{figure}[H]
\begin{center}
\includegraphics[width=0.46\textwidth]{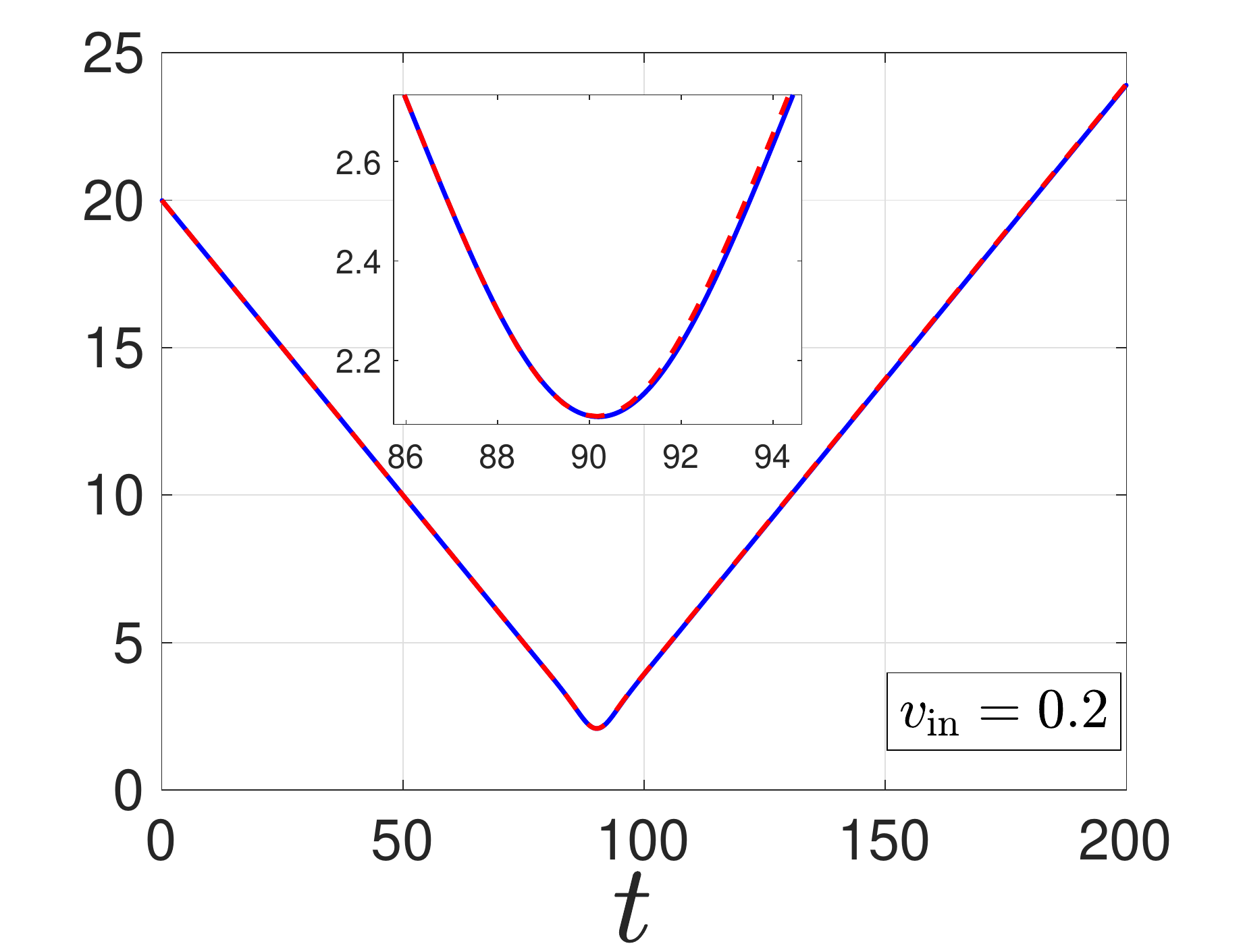}
\includegraphics[width=0.46\textwidth]{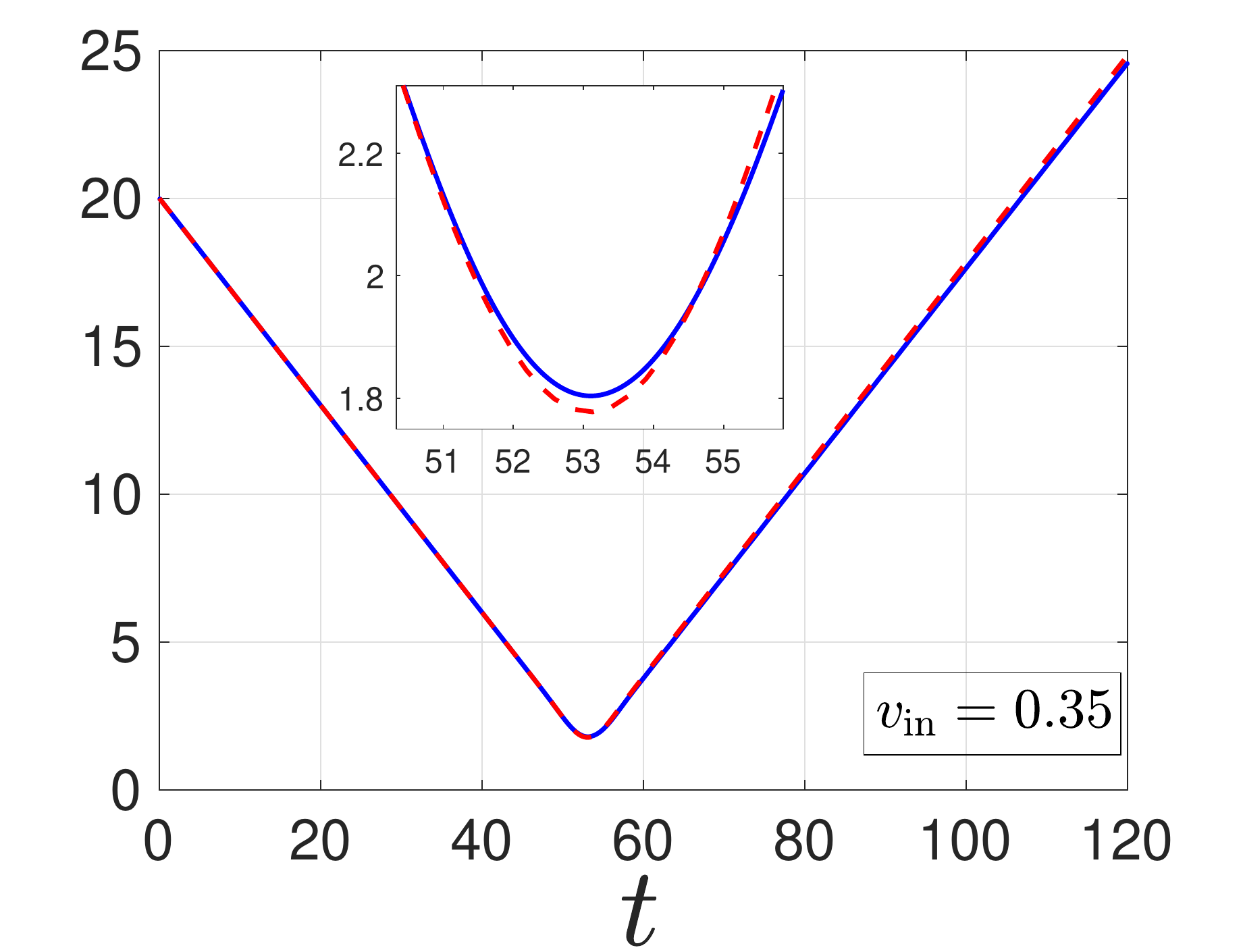}
\includegraphics[width=0.46\textwidth]{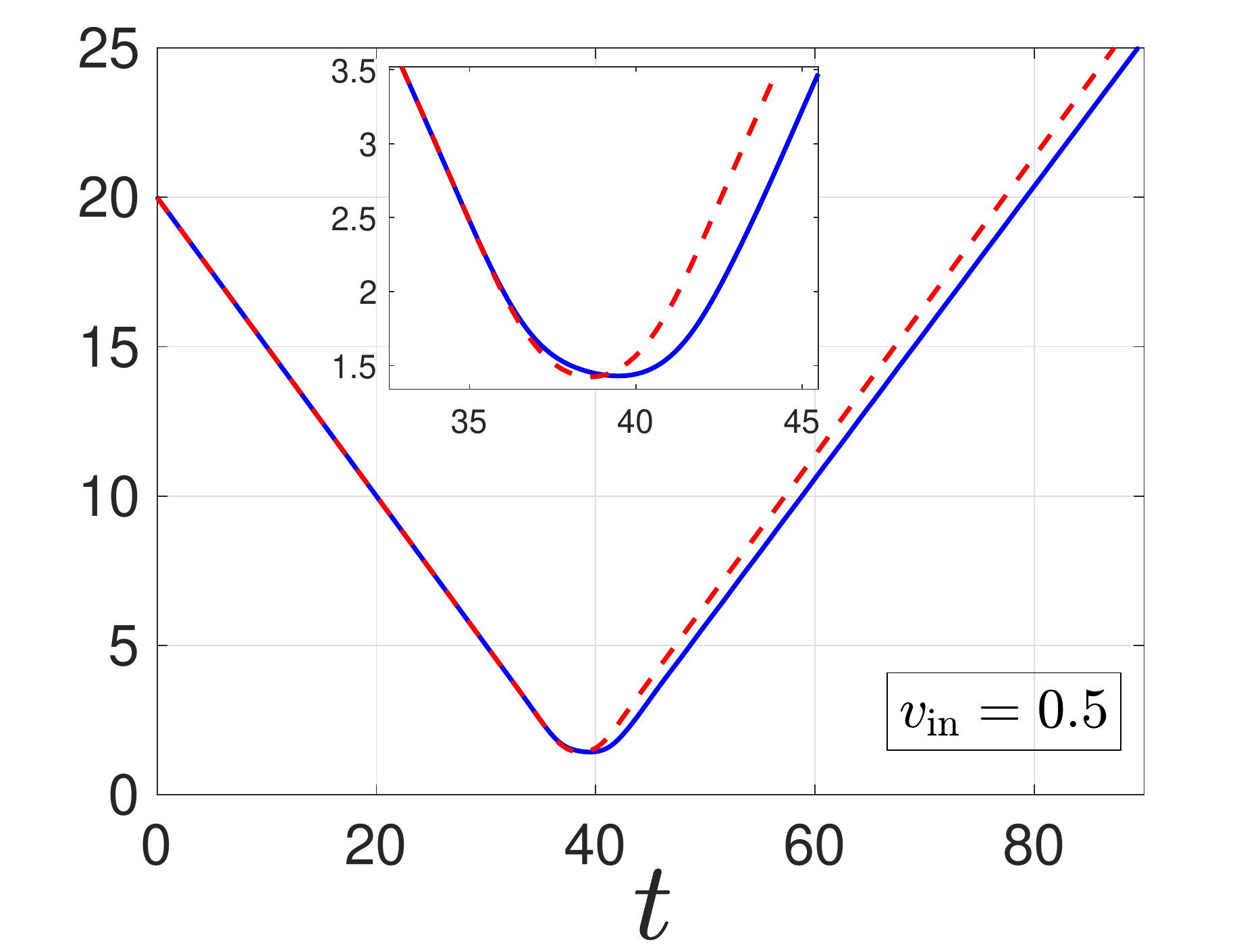}
\includegraphics[width=0.46\textwidth]{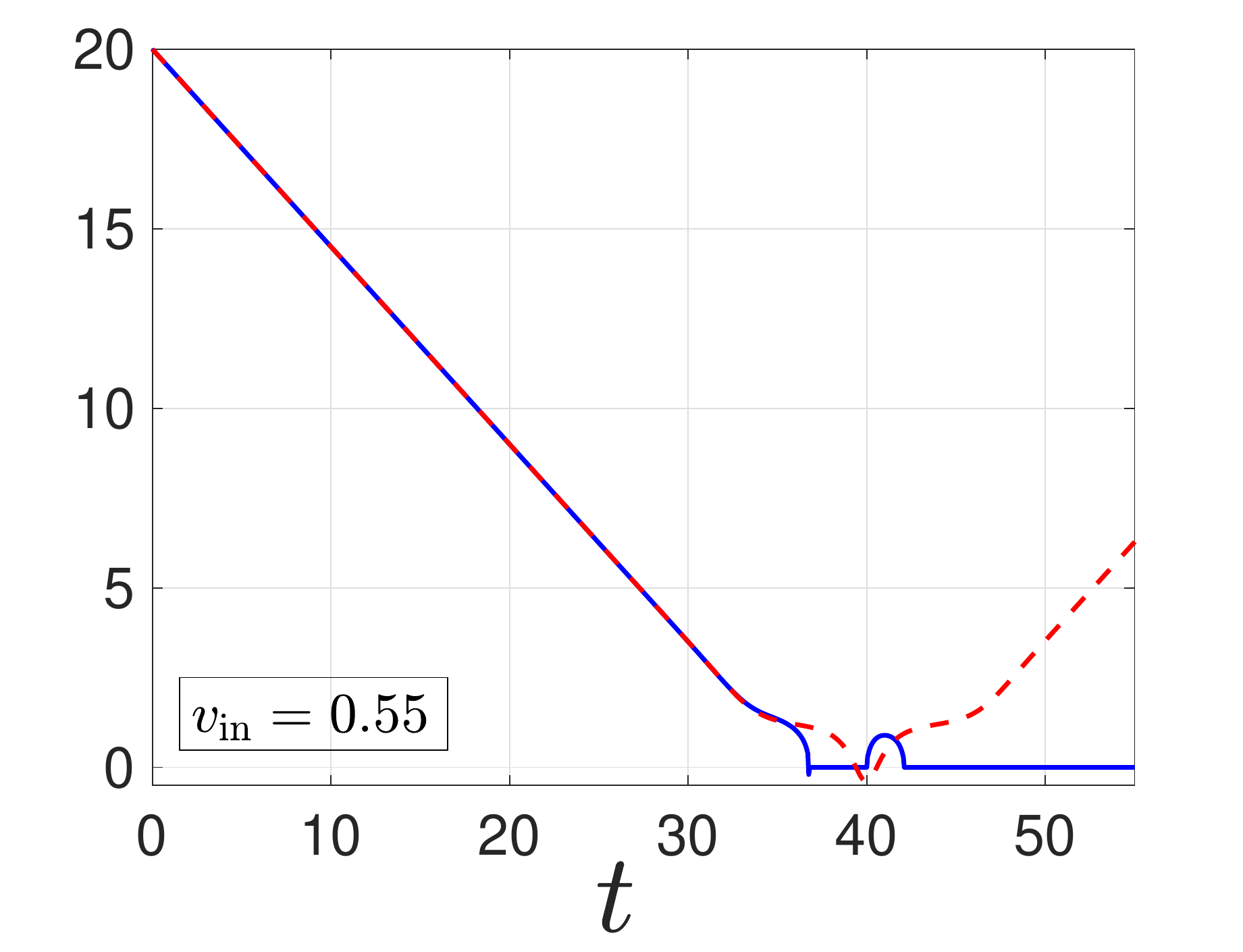}
\includegraphics[width=0.46\textwidth]{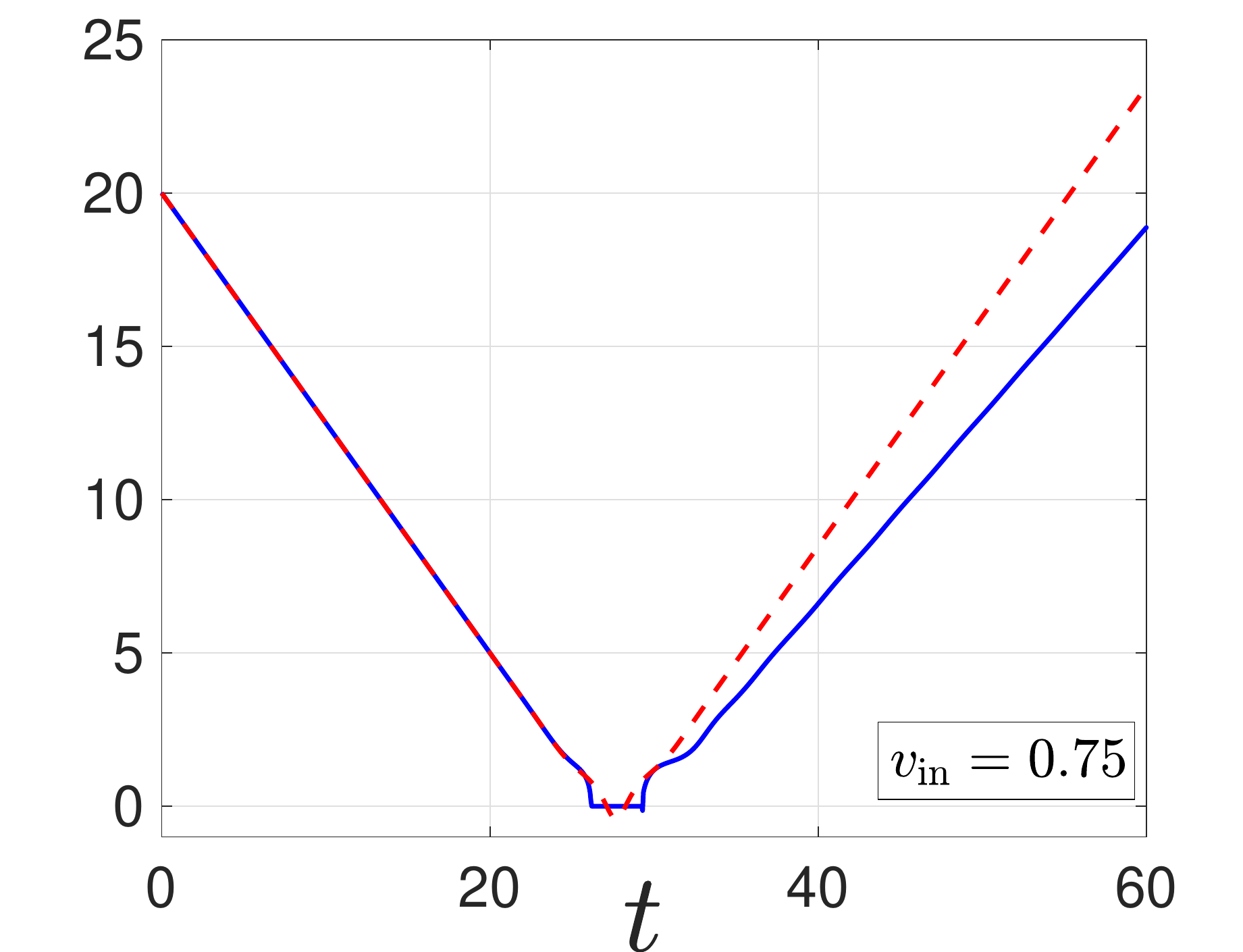}
\end{center}
\caption{The figure shows the ODE solution $X(t)$ (in dash red line) on top of PDE solution (in solid blue curve) for various values of $v_{\mathrm{in}}$. For small values $v_{\mathrm{in}}$, we observe a good match, but as $v_{\mathrm{in}}$ increases, a divergence occurs after the collision. }
\label{cc_simple}
\end{figure}

In Figure \ref{cc_simple}, we plot the PDE  and the ODE solutions
(obtained using CC method) for various values of
$v_{\mathrm{in}}$. The PDE plot in the figure is the position of the
approximate center of the antikink solution as defined by its
intersection with the x-axis. As seen in the figure, we get a nearly perfect match for $v_{\mathrm{in}}=0.2$. When we increase $v_{\mathrm{in}}$ to $0.35$, we see a slight difference. That difference gets more noticeable when $v_{\mathrm{in}}$ is $0.5$.  When we take $v_{\mathrm{in}}=0.55$, we see a divergence after the collision. For $v_{\mathrm{in}}=0.75$, we only see a good match until the collision.

%
%

\section{Conclusions and Future Work}

In the present work we have explored the biharmonic $\phi^4$
(B$\phi^4$)
model and
some of the central properties of its kink solutions. We have
illustrated
that the model has kinks with tails that are distinctly different than
those
of the standard $\phi^4$ model in that they bear an oscillatory
structure (instead of the monotonic kinks in $\phi^4$). We have
also performed a spectral analysis of both static and traveling
kinks. The case of the latter is not as straightforwardly mappable
to the former in the B$\phi^4$ model due to the absence of the Lorentz
invariance. Both static kinks and traveling ones below a certain speed
appear to have an internal mode in the B$\phi^4$ model. Lastly,
we tackled collisions between a kink and an antikink. These were found
to
be quite different than the complex fractal collision structure of
the regular $\phi^4$ model. Here, the scenarios turned out to be far
more clear in their structure with elastic apparent repulsion between
the wave occurring at small speeds, collision into an infinite bounce
capture for a short range of intermediate ones and eventually
inelastic
single bounces at large speeds. Nevertheless, a different source of
complexity was unveiled in the two transition regions between these
three regimes. Namely, a delicate oscillatory logarithmic dependence of the
outgoing
vs. incoming velocity was revealed that was intuitively attributed to
the
more complex tail and associated interaction structure of the two
waves,
but which also merits further elaboration in future work.

Lastly, we attempted the most simple version of the CC method
towards characterizing the B$\phi^4$ model kink-antikink collisions. 
The CC method we have applied  has only one degree of freedom, so it
is expected not to fully capture the PDE behavior.
It is natural to expand this by attempting to take into consideration
the internal mode of the kink and antikink. 
Then, the corresponding ansatz that is relevant to consider becomes:
\begin{equation}\label{kink-antikink_ans}
u (x,t)=\varphi_0(x+X(t))-\varphi_0(x-X(t))-1+A(t)(S(x+X(t))-S(x-X(t))) 
\end{equation}
where $X(t)$ which is the time-dependent displacement of the kink from
the origin and $A(t)$ is the amplitude of the internal mode
perturbation and $S(x)$ is the eigenfunction corresponding to the lowest (positive)
eigenfrequency of the kink. A question that arises, however, in this
setting is which frequency it is suitable to consider, as the static
and traveling kink are not effectively equivalent and there is a
dependence
of the internal mode frequency on the corresponding speed.
Using, as is done in $\phi^4$ the frequency of the static kink and
attempting to solve the corresponding ODE system, one obtains
a numerical instability around $X=0$. This has been a common issue
with
$\phi^n$-models, as discussed, e.g., in the recent review
of~\cite{cuevas}: it
has been dubbed the null-vector problem~\cite{caputo}. Reduced ODE
systems were studied in the earlier works assuming  the terms with
higher order derivatives of  $A(t)$ and $X(t)$ stayed negligible. Our
numerical computations suggest that this is not a suitable assumption around
$X=0$. Hence, clearly there are some important challenges ahead,
especially
as regards an understanding of the phenomenology of collisions
and, more generally, of kink-antikink interactions and ``bound
states''.
These appear to us to certainly be worthwhile to consider in future
studies
and will accordingly be reported in future publications.


\begin{thebibliography}{100}


\bibitem{levandosky}
     \newblock S. Levandosky,  
     \newblock Stability and instability of fourth order solitary
     waves,  \newblock \emph{ J. Dynam. Differential
Equations}, \textbf{10}, 151
     (1998).
    

\bibitem{champneys}
     \newblock  A.R. Champneys, P.J. McKenna and P.A. Zegeling,  
     \newblock Solitary waves in nonlinear beam equations: stability, fission and fusion,
     \newblock \emph{ Nonlinear Dynamics}, \textbf{21}, 31
      (2000).
     
 \bibitem{CM}
     \newblock  Y. Chen, P.J. McKenna,  
     \newblock Traveling waves in a nonlinearly suspended beam:
     theoretical results and numerical observations,     \newblock
     \emph{ J. Differential Equations}, \textbf{136},
     325
     -355
     (1997).
     
\bibitem{karageorgis}
     \newblock  P. Karageorgis, P. J. McKenna,  
     \newblock The existence of ground states for fourth-order wave equations, 
     \newblock \emph{ Nonlinear Anal.}, \textbf{73}, 367
      (2010).

    \bibitem{pqs} A. Blanco-Redondo, C. Martijn de Sterke, J.E. Sipe,
      T.F. Krauss,
      B.J. Eggleton and C.  Husko 
      \newblock Pure-quartic solitons,
      \newblock \emph{Nature Comms.} {\bf 7}, 10427 (2016).

    \bibitem{atanas} I. Posukhovskyi and A. Stefanov,
      \newblock On the normalized ground states for the Kawahara
      equation and a fourth order NLS,
      arXiv:1711.00367.

    \bibitem{ablowitz} M.J. Ablowitz,
      \newblock Nonlinear Dispersive Waves,
      \newblock Cambridge University Press (Cambridge, 2011).

\bibitem{beam_demirkaya} A. Demirkaya, M. Stanislavova, Numerical results on existence and stability of standing
  and traveling waves for the fourth order beam equation,
  \emph{Discrete Contin. Dyn. Syst.-B},
   {\bf 24},  197 (2019).

\bibitem{Moon} F. C. Moon and P. J. Holmes, A magnetoelastic strange attractor, \emph{Journal of Sound and Vibration},  {\bf 65}, 275 (1979).

\bibitem{Campbell} D. K. Campbell, J. S. Schonfeld, and C. A. Wingate, Resonance structure in kink-antikink interactions
in $\phi^4$ theory, \emph{Phys. D}, {\bf 9}, 1 (1983).

\bibitem{Belova} T.I. Belova and A.E. Kudryavtsev,
Solitons and their interactions in classical field theory, 
  \emph{Phys. Usp.}, {\bf  40}, 359 (1997).

\bibitem{cuevas} P.G. Kevrekidis, J. Cuevas-Maraver (Eds.),
  A Dynamical Perspective on the $\phi^4$ model,
 Springer-Verlag (Heidelberg,  2019).

\bibitem{Ann} P. Anninos, S. Oliveira, and R.A. Matzner,
  Fractal structure in the scalar $\lambda (\phi^2-1)^2$ theory,
 \emph{Phys. Rev. D}, {\bf 44}, 1147 (1991).

\bibitem{goodman} R.H. Goodman, R. Haberman,
  \newblock Kink-antikink collisions in the phi-four equation: The
  n-bounce resonance and the separatrix map,
  \newblock \emph{SIAM J. Appl. Dyn. Sys.}, {\bf 4}, 1105
  (2005).

\bibitem{goodman2} R.H. Goodman,
   	Chaotic scattering in solitary wave interactions: A singular
        iterated-map description, 	\emph{Chaos},  {\bf 18},
        023113 (2008). 
   
\bibitem{weigel} H.~Weigel,
  Kink--antikink scattering in $\phi^4$ and $\phi^6$ models,
  \emph{J. Phys. Conf. Ser.},  {\bf 482}, 012045 (2014), arXiv:1309.6607.

\bibitem{weigel2} I.~Takyi and H.~Weigel,
  Collective coordinates in one-dimensional soliton models revisited,
\emph{Phys.\ Rev.\ D}, {\bf 94}, 085008 (2016), arXiv:1609.06833.
 
 
  
  
 \bibitem{trefethen} L. N. Trefethen,
      \newblock Spectral Methods in MATLAB,
      \newblock SIAM (Philadelphia, 2000).

  


\bibitem{christov}  I. C. Christov, R. Decker, A. Demirkaya,
  P. G. Kevrekidis, V. A. Gani, Long range interactions,
  \emph{Phys. Rev. D}, {\bf 99}, 016010, (2019). 



   
\bibitem{Sugiyama} T. Sugiyama,
Kink-antikink collisions in the two-dimensional $\phi^4$ model,
  \emph{Prog. Theor. Phys.},  {\bf 61}, 1550 (1979).

\bibitem{caputo} J.G. Caputo and N. Flytzanis,
  Kink-antikink collisions in sine-Gordon and 
$\phi^4$ models: Problems in the variational approach,
  \emph{Phys. Rev. A}, {\bf 44}, 6219 (1991).
  




      


\end{thebibliography}
\end{document}